\documentclass[journal]{vgtc}              

\graphicspath{{figs/}} 
\usepackage[dvipsnames,svgnames]{xcolor}
\usepackage{amsfonts}
\usepackage{amsmath}
\usepackage{array}
\usepackage{subcaption} 
\usepackage{todonotes}
\usepackage{setspace}

\definecolor{sandybrown}{HTML}{c56632}
\definecolor{neptune}{HTML}{00724a}
\definecolor{poloblue}{HTML}{2853c9}
\definecolor{shocking}{HTML}{b2006d}

\newcommand{\para}[1]{\vspace{2pt}\noindent\textbf{#1}}
\newcommand{\challenge}[1]{{\color{blue}\hyperref[challenge-#1]{\textbf{#1}}}}
\newcommand{\challengedef}[1]{{\color{blue} \phantomsection \label{challenge-#1} \textbf{#1}}}

\newcommand{\Rom}[1]{\uppercase\expandafter{\romannumeral #1\relax}}
\newcommand{\rom}[1]{\expandafter{\romannumeral #1\relax}}

\newcommand{\heatmapraw}{glyphs/heatmap.png}
\newcommand{\jaccardraw}{glyphs/jaccard.png}
\newcommand{\scatterplotraw}{glyphs/scatter.png}
\newcommand{\datasetraw}{glyphs/dataset.png}

\newcommand{\heatmap}{\raisebox{-0.6ex}{\includegraphics[height=1.1em]{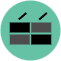}}}
\newcommand{\jaccard}{\raisebox{-0.6ex}{\includegraphics[height=1.1em]{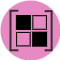}}}
\newcommand{\scatterplot}{\raisebox{-0.6ex}{\includegraphics[height=1.1em]{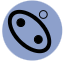}}}
\newcommand{\dataset}{\raisebox{-0.6ex}{\includegraphics[height=1.1em]{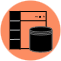}}}

\newcommand{\datasetcolor}[1]{\textcolor{sandybrown}{#1}}
\newcommand{\scattercolor}[1]{\textcolor{poloblue}{#1}}
\newcommand{\jaccardcolor}[1]{\textcolor{shocking}{#1}}
\newcommand{\heatmapcolor}[1]{\textcolor{neptune}{#1}}

\newcommand{\toolname}{\textit{ChannelExplorer}}
\onlineid{1442}
\vgtccategory{Research}
\vgtcpapertype{Analytics and Decisions}

\title{{\toolname}: Exploring Class Separability\\ Through Activation Channel Visualization}

\author{
  \authororcid{Md Rahat-uz-Zaman}{0000-0001-6728-7569},
  \authororcid{Bei Wang}{0000-0002-9240-0700}, and
  \authororcid{Paul Rosen}{0000-0002-0873-9518}
}
\authorfooter{
    \item Md Rahat-uz-Zaman is with the University of Utah. E-mail: rahatzamancse@gmail.com.
    \item Bei Wang is with the University of Utah. E-mail: beiwang@sci.utah.edu.
    \item Paul Rosen is with the University of Utah. E-mail: paul.rosen@utah.edu. 
}

\abstract{%
Deep neural networks (DNNs) achieve state-of-the-art performance in many vision tasks, yet understanding their internal behavior remains challenging—particularly how different layers and activation channels contribute to class separability. 
We introduce {\toolname}, an interactive visual analytics tool for analyzing image-based outputs across model layers, emphasizing data-driven insights over architecture analysis for exploring class separability.
{\toolname} summarizes activations across layers and visualizes them using three primary coordinated views: a Scatterplot View to reveal inter- and intra-class confusion, a Jaccard Similarity View to quantify activation overlap, and a Heatmap View to inspect activation channel patterns.
Our technique supports diverse model architectures, including CNNs, GANs, ResNet and Stable Diffusion models. 
We demonstrate the capabilities of {\toolname} through four use-case scenarios: 
(1) generating class hierarchy in ImageNet, 
(2) finding mislabeled images, 
(3) identifying activation channel contributions, and
(4) locating latent states' position in Stable Diffusion model.
Finally, we evaluate the tool with expert users.
}


\teaser{
  \centering
  \includegraphics[width=\linewidth, alt={The overview of the ChannelExplorer WebView.}]{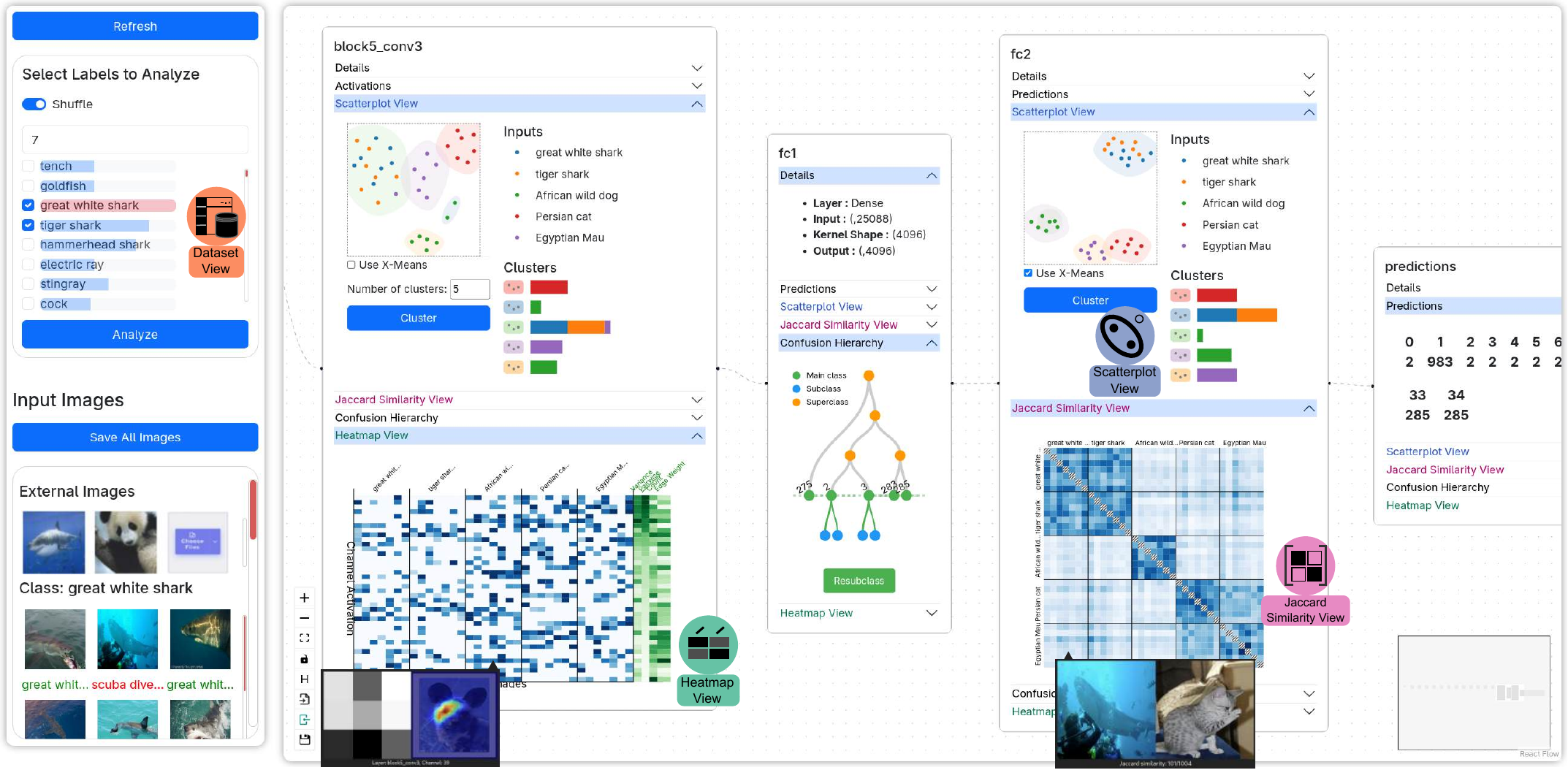}
  \caption{%
\toolname{} has 7 images loaded from 5 classes using \datasetcolor{Dataset View} \dataset{}.
Each layer is a node in the model's node-link diagram view. The tool has three key views: 
(\rom{1}) \scattercolor{Scatterplot View} \scatterplot: 2D embeddings of activation summaries highlight class confusion patterns—e.g., overlapping clusters for shark classes (inter-class confusion) and multiple clusters within the dog class (intra-class confusion); 
(\rom{2}) \jaccardcolor{Jaccard Similarity View} \jaccard: a similarity matrix showing activation overlap between inputs—e.g., high similarity between shark classes and cat classes; 
(\rom{3}) \heatmapcolor{Heatmap View} \heatmap: a heatmap of activation channel summaries, enabling identification of channels that respond exclusively to certain classes and may encode class-distinctive features-e.g., activation channel for dogs' eyes.
The confusion hierarchy view visualizes a class hierarchy from the layer's class confusion.
}
  \label{fig:teaser}
}

\begin{document}
\maketitle

\setstretch{0.985}
\section{Introduction}
\label{sec:introduction}

Deep Neural Networks (DNNs) are widely used in machine learning (ML) tasks, including image classification~\cite{KrizhevskySutskeverHinton2017}, object detection~\cite{GirshickDonahueDarrell2014}, semantic segmentation~\cite{SultanaSufianDutta2020}, and facial recognition~\cite{HuYangYi2015}. 
Understanding hyperparameters, model architecture, and various metrics such as loss, accuracy, and ROC curves---along with associated visualization tools~\cite{FraternaliMilaniTorres2022, GranziolWanGaripov2020, HoroiHuangWolf2021, BainTokarevKothari2021}---can aid in building better models~\cite{YuZhu2020}. However, interpreting and debugging what the model perceives from the input data---encoded in the neural network activations from hidden layers---remains a significant and persistent challenge.

In this paper, we examine neural network activations to enhance the interpretability of DNNs that contain image-based layers. An \emph{image-based layer} in a neural network is a layer that processes data in the form of image-like tensors. Common types of image-based layers include convolutional layers, deconvolution layers, pooling layers, affine transformation layers, upsampling layers, and residual blocks (i.e.,~composition of image-based layers). We introduce a visual analytic tool called {\toolname} that assists ML experts in exploring, refining, and understanding class separability for model design, diagnostics, and improvement. {\toolname} is applicable to a wide variety of model architectures that contain image-based layers, such as Convolutional Neural Networks (CNNs), U-Net, Generative Adversarial Networks (GANs), Stable Diffusion models (SD), Fully Convolutional Networks (FCNs), ResNet (Residual Networks), and DenseNet. 

The importance of a model's ability to perform class separability spans from classification to many non-classification vision tasks such as object detection, segmentation, and image generation. For instance, object localization models like YOLO~\cite{yolov3} first identify detection ranges and then \emph{classify} those objects. Also, image describing models like CLIP~\cite{DBLP:journals/corr/abs-2103-00020} align text and image embeddings through \emph{classification} tasks (e.g., identifying which image matches a text description). 
Despite being a high-accuracy model, class ambiguity can lead to inconsistent outputs, especially in cases where prediction confidence is low. It is imperative to understand where current output falls within the overall prediction distribution to evaluate its quality. This is especially critical when the model's decisions affect areas like safety, healthcare, or finance. 

Exploring class separability ensures that the model can effectively distinguish between classes, leading to improved (1)~interpretability, (2)~performance, (3)~generalization and robustness, as well as (4)~enhanced data quality, and (5)~guided model refinement. By examining and enhancing class separability, ML experts can detect issues, refine model architecture, and make well-informed decisions throughout the design and training process. 
(1)~\emph{Interpretability}: Studying class separability provides insights into how the model processes data and whether it is learning meaningful features for classification. 
(2)~\emph{Performance}: Since both the model architecture and training data jointly determine accuracy and speed, it is essential to evaluate performance in terms of data separability~\cite{XueZhangJiang2023}. 
A well-separated feature space leads to distinct decision boundaries, increasing the likelihood of accurate predictions. Understanding class separability helps pinpoint cases where certain classes are difficult to distinguish, allowing for targeted improvements to enhance accuracy.
(3)~\emph{Generalization and Robustness}: Both in classification and non-classification tasks, data group separability helps explain how well a model generalizes, and studies have shown that CNN layers improve linear separability while determining which information is retained or discarded, influencing generalization~\cite{BelcherPrugel-BennettDasmahapatra2020}.
A well-separated class space ensures the model is less sensitive to noise, outliers, or small perturbations in the input, contributing to both robustness and adaptability across diverse test cases.
(4)~\emph{Data Quality}: Analyzing class separability can help detect poorly separated classes caused by mislabeling, insufficient data (such as class imbalance~\cite{GhoshBellingerCorizzo2024}), and inconsistencies in annotations.
(5)~\emph{Model Refinement}: Evaluating separability assists in designing more efficient models that enhance the separation of challenging-to-classify classes~\cite{LuoWongKankanhalli2020} or penalize ambiguous regions in the feature space~\cite{BilalJourablooYe2017, AlsallakhJourablooYe2018}.

We present {\toolname}, a tool designed to investigate class separability through channel activations, facilitating all five types of model improvement discussed above. Our contributions are three-fold. 

\begin{itemize}[noitemsep,leftmargin=*]

\item \textbf{Enhanced interpretability.} {\toolname} highlights class separability within the activation space at each layer, making the model’s decision-making process more transparent and~\emph{interpretable}. The combination of \scattercolor{Scatterplot View} and \jaccardcolor{Jaccard Similarity View} reveals both inter-class confusion (when similar neuron activations between classes cause confusion) and intra-class confusion (when variation in activations within a class complicate classification). 
It also helps ML experts derive a class hierarchy that pinpoints weaknesses in the model surrounding poor class separability. 

\item \textbf{Guided model diagnostics.}  {\toolname} facilitates guided drill-down exploration from whole model to individual activation channels through three views: \scattercolor{Scatterplot View}, \jaccardcolor{Jaccard Similarity View} and \heatmapcolor{Activation Heatmap View}. By employing an overview-first, details-on-demand approach, it visualizes layers' and activation channels' contributions, highlighting their \emph{performance} impacts and \emph{generalizability}. This helps ML experts identify areas where the model struggles, and opportunities for \emph{model refinement}. 
\item \textbf{Enriched data quality.} {\toolname} helps identify mislabeling and annotation errors. Poorly separated classes identified via the \scattercolor{Scatterplot View} and the \heatmapcolor{Heatmap View} often indicate mislabeling or inconsistent annotations.  {\toolname} helps in detecting mislabeled samples for correction, leading to better \emph{data quality}.  
\end{itemize}

Finally, we offer an open-source implementation of {\toolname}, allowing ML experts to analyze class separability across different model-data combinations.

\section{An Introduction to Neural Network Activations}
\label{sec:background}

A Deep Neural Network (DNN) consists of multiple layers of neurons as processing units: the \emph{input layer} receives raw data (e.g., an image), the \emph{hidden layers} analyze the data by detecting patterns, and the \emph{output layer} generates the final result (e.g., a classification label such as \emph{goldfish} or \emph{shark}). 
However, the interpretability of a neural network lies in its hidden layers, where the network discovers a new, hidden representation of the input at each layer. 

\begin{figure}[!ht]
    \centering
    \includegraphics[width=\linewidth]{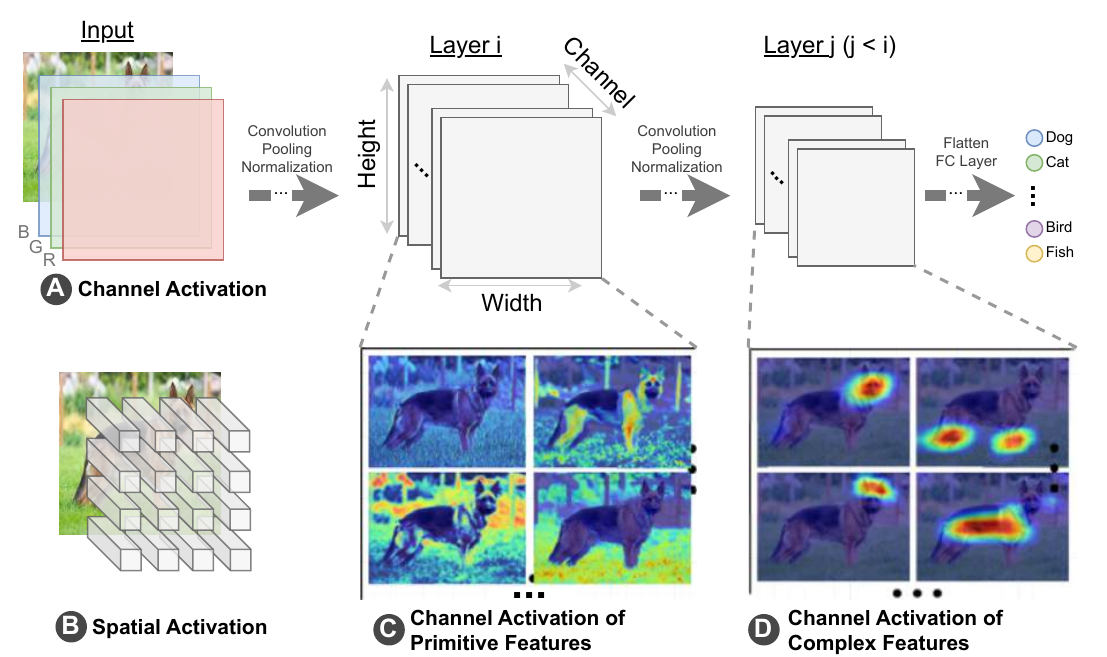}
    \caption{(A) and (B) show two ways for image activation analysis. We use (A) as each channel represents the presence of features in the input. Channel activations of two layers are shown in (C) and (D). Layer $i$ identifies basic features (i.e., grass texture, fur texture, edges, etc.), whereas layer $j$ identifies complex features (i.e., dog's head, body, etc.).}
    \label{fig:basic-advance-features}
\end{figure}

The learned representation of images at each layer is a tensor and can be visualized as a 3D cube, where each cell represents an activation, that is, the extent to which a neuron fires. The x- and y-axes represent the spatial positions of pixels in the image, while the z-axis denotes the channels. Slicing such a cube in x- and y-axis produces vectors called \emph{spatial activations}, and slicing through z-axis produces matrices called \emph{channel activations}~\cite{OlahSatyanarayanJohnson2018}; as illustrated in \cref{fig:basic-advance-features}(A and B). 


Studies have shown that the channel activations of each layer convey information about the features the layer is attempting to detect~\cite{hohman2020summit, DBLP:journals/corr/TzengHZSD14, DBLP:journals/corr/BauZKOT17, carter2019activation, dumitruVisualizing2009, DBLP:journals/corr/YosinskiCNFL15}. Earlier layers typically extract the basic features (e.g., edges, gradients, intensity variations, etc.), whereas latter layers extract more complex features that support the detection of objects and patterns~\cite{olah2017feature}. \cref{fig:basic-advance-features} shows that channels in early layer~$i$ activate sparsely for basic textures, whereas channels in later layer~$j$ display a more focused region in the image, detecting complex features. 


Our tool {\toolname} focuses on exploring the channel activations for enhancing the interpretability of DNNs that contain image-based layers. 
\section{Related Works}
\label{sec:related_works}

We divide existing works related to class confusion analysis, class hierarchy generation, and DNN visualization tools into 7 categories. 

\para{Visualizing model architecture, metrics and hyperparameters.}
Many deep learning visualization tools focus on model parameters and evaluation metrics like precision, recall, F1-score and confusion matrix. 
Tools such as ConfusionFlow~\cite{confusionflow}, InstanceFlow~\cite{instanceflow}, CNNComparator~\cite{cnncomparator}, and EnsembleMatrix~\cite{ensemblematrix} visualize these metrics and their changes across epochs or model ensembles. 
Many model visualization systems like DQNViz~\cite{dqnviz}, ReVACNN~\cite{revacnn}, DeepTracker~\cite{deeptracker}, ManiMatrix~\cite{manimatrix}, RadViz~\cite{radviz}, Seifert et al.~\cite{seifert_visualizations_2017}, Alsallakh et al.~\cite{histogramclassviz}, and Squares~\cite{ren2016squares} provide similar visualizations of model weights and confusion matrices—either at the class level or per instance during training or inference. 
Neo~\cite{neo_conf_matrix} enhances this by treating the confusion matrix as a probabilistic distribution and creating an interactive class hierarchy.
However, relying only on scores and confusion matrices can not identify generalizability for all classes. For instance, models can have accurate predictions with low confidence or overlapping activation patterns with other classes, which cannot be captured with accuracies or confusion matrices.
Finally, these tools largely overlook the internal activation dynamics across layers, which are critical for model transparency and interpretability.  

\para{Visualization of Raw Activation Channels}. 
Some visualization tools directly normalize and show the activation channels. DeepVis~\cite{DBLP:journals/corr/YosinskiCNFL15}, Tensorflow Playground~\cite{DBLP:journals/corr/abs-2101-04141} and Quiver~\cite{bianquiver} show the channels one layer at a time. 
However, they lack the visualizations of analytics on the activation channels.

\para{Activation Channels as Features of a Model}. 
Several studies propose that activation channels can be used as interpretable features in another machine-learning model to determine feature relevance~\cite{DBLP:journals/corr/TzengHZSD14}.
To interpret layer activations, the Network Dissection framework applied pixel-wise segmentation and measured the alignment between different activation channels~\cite{DBLP:journals/corr/BauZKOT17}. 
Extending this work, Net2Vec computed the Intersection over Union (IoU) score between two classes (e.g., dog and airplane) to interpret the feature similarity~\cite{DBLP:journals/corr/abs-1801-03454}.
TCAV used a different approach, applying binary classification to the activations of each layer to identify the patterns (e.g., zig-zag pattern) that are relevant to a specific class (e.g., zebra)~\cite{pmlr-v80-kim18d}.
ACE clusters the latent state of segments of images and use TCAV to apply concept importance score to each segment patterns~\cite{ace_ghorbani}. 
ConceptExplainer uses both TCAV and ACE to create an open-source visualization tool to explore concept-based explanations for non-expert users~\cite{conceptexplainer}.
Finally, ActiVis showed an interactive table view, where columns are average neuron activations in a layer, and rows are classes~\cite{DBLP:journals/corr/KahngAKC17}. 
These methods provide useful insights, but they lack visualization at the granularity of channels or input samples, which is important to identify channels that do not help in the target task.

\para{Dimension Reduction on Activation Channels}. 
Some techniques reduce the dimension of activations and project them into 2D or 3D space.
Activation Atlas~\cite{carter2019activation} and the t-SNE visualization of CNNs~\cite{AndrejT-SNE} collected random spatial activations and visualized them in a 2D plane using UMAP~\cite{mcinnes_umap_2020} and t-SNE~\cite{maaten_visualizing_2008}, respectively. Each image in the 2D embedding represents the average activations of the model. 
Pezzotti et al. used HSNE~\cite{HSNE_Pezzotti} embeddings to analyze and show stable layers during training~\cite{pezzotti2017deepeyes}.  
Since most of these studies focus on average class activations, the embeddings cannot visualize individual images or layers for debugging the model.

\para{Layer Activation Maximization}. 
Some neural network interpretation technique involves maximizing the activations of a layer by iteratively updating an input noise that turns into a feature interpretable image~\cite{dumitruVisualizing2009, DBLP:journals/corr/YosinskiCNFL15}.
Erhan et al.\ performed this using Stacked Denoising Autoencoders to show that reasonable qualitative interpretations of high-level features are possible~\cite{dumitruVisualizing2009}. 
Plug \& Play Generative Networks improved these approaches using a Conditional GAN that can produce higher-quality diverse images~\cite{DBLP:journals/corr/NguyenYBDC16}. 
Instead of noise, Inceptionism modified an image (e.g., sky) using gradient ascent algorithms to fool a CNN model to predict it as a different class (e.g., dog, snail, fish, etc.)~\cite{google_inceptionism_2015}. 
OpenAI Microscope~\cite{openai_microscope} identified the highest activated neurons for inputs and visualized the features of those layers using DeepDream~\cite{google_inceptionism_2015}. 
These visualizations are mostly used for educational or interpretation purposes, are limited to CNN models, and cannot help directly identify errors or improvements in the model.

\para{Aggregation, Selection, Filter, and Summarization of Activation Channels}. 
Liu et al.\ selected activation layers, channels, or neurons based on hierarchical feature recognition capability of layers (e.g., an earlier layer failed to detect the ear of a panda, so later layers failed to detect that panda) using a datapath visualization method~\cite{DBLP:journals/corr/abs-1810-03913}.
Olah et al.\ made this selection for layers that fire the highest activations and generated feature-enhancing images for those layers only~\cite{olah2017feature}.
Goh et al.\ analyzed more low-level (activation channels) and grouped them at each layer by their detected features (e.g., color, brand, typography, founding, emotion neurons)~\cite{goh2021multimodal}. Finally, Hohman et al. analyzed the number of activated neurons for different classes and created activation pathways for visualizing feature identification~\cite{hohman2020summit} and adversarial attacks~\cite{das2020bluff}. 
These visualizations have limited scope because many features identified by layers or channels are not explainable or have mental concepts. Molnar showed that 193 out of 512 channels in a layer of the VGG model are not interpretable~\cite{molnar2022}. 

\para{Model Explainers for Educational Purpose}. 
Tools such as CNN Explainer~\cite{Wang2020CNNEL}, Transformer Explainer~\cite{cho2024transformerexplainerinteractivelearning}, Diffusion Explainer~\cite{lee2024diffusionexplainervisualexplanation}, and GAN Lab~\cite{Kahng2018GANLU} offer intuitive web-based interfaces aimed at educating beginners on the functionality of these models. However, their approaches are designed exclusively for educational purposes and lack utility for tasks such as model debugging or optimization.

\subsection{Comparison with Existing Works}
\label{sec:comparison_exisiting}

The area of DNN explainability is rich. While \toolname{} shares many qualities with prior works, it also provides some unique characteristics.

\para{Architecture Overview.}
Most traditional visualizations represent the overall DNN using node-link diagrams and depict image-based layers in overviews with stacked images—typically displaying only a handful of activation channels (3–5 out of thousands)~\cite{bellgardt_immersive_2020, harley_interactive_2015, samuel_visualizing_2018}.
Some tools show layer details separately~\cite{openai_microscope, schorr_neuroscope_2021, hohman2020summit, rathore_topoact_2021}, limiting simultaneous multi-layer analysis.
In contrast, \toolname{} uses a “Details Inside” design with collapsible panels embedded in each layer node, enabling scalable, side-by-side comparisons while preserving the network's mental model.

\para{Improved Class Hierarchy Generation.}
Certain prior works are similar to isolated components of our visualization methodology, such as dimensionality reduction~\cite{AndrejT-SNE,pezzotti2017deepeyes} or confusion matrix visualization~\cite{confusionflow,instanceflow}. 
These approaches show the interpretability of neural networks but fall short of enabling error detection across both data and model components at different levels of granularity (layers, channels).
For instance, Alsallakh et al.~\cite{AlsallakhJourablooYe2018} in the Blocks system enabled users to construct a class hierarchy of only super-classes from a confusion matrix, visualize activation similarity between classes, and exploit the class hierarchy to force layers to classify a fixed number of classes. 
While this provides useful insights and model improvement, our observations using \toolname{} reveal that many modern larger deep networks do not perform classification in earlier layers. Instead, these layers identify basic features that serve as inputs for downstream layers in any task. This renders tools, such as Blocks, not applicable as they assume early layers perform classification. 
\toolname{} addresses this by analyzing class separability in terms of pseudo-classes (user-defined groups of selected images as input), enabling applicability for both classification and non-classification tasks. Unlike existing works like Blocks~\cite{AlsallakhJourablooYe2018}, 
our visualization generates a class hierarchy based on each layer and supports both sub-classification and super-classification. 
The earlier class hierarchy is based on the presence of common primitive features, and the later layers' class hierarchy is based on task-driven class separability.

\para{Toward Multi-level Contextual Analysis.}
All works discussed in~\cref{sec:related_works} help explain neural networks but lack the ability to locate areas of interest in a model or dataset during visualization. 
We address this limitation by using a contextual hierarchical analysis process, where users can explore model behavior at multiple levels of abstraction--from overall class-level confusion across the model, to confusion within specific layers, down to the contribution of individual activation channels.  

\para{Transferable Model-Agnostic Debugging.}
Furthermore, these works are tailored to specific models and lack transferability to custom architectures or datasets.
\toolname{} focuses on the \emph{debuggability} of the model and dataset to identify limitations in specific regions and is designed to work with any image-based model architecture. 


In sum, while visualizing metrics and raw activations per layer is not new, \toolname{} extends this by introducing ordered summarization, class-driven abstraction, and channel-aware exploration--enabling the identification of class ambiguity, discovery of latent class hierarchies, and interpretation of channel contributions--capabilities largely missing in prior work.
It is a general-purpose, data-driven system built to support both interpretation and debugging of complex, modern neural networks.

\section{Design Overview}
\label{sec:design_overview}
We aim to develop an interactive visualization to help people apply visual analytics on any image-based model and dataset. 


\subsection{Visualization Challenges}
\label{sec:challenges}
From our evaluation of prior work and discussions with various deep learning and visualization experts, we identified three key visualization challenges associated with class separability. 


\para{\challengedef{C1}. Difficulty visualizing the abundance of activation channels.} 
Visualizing important activation channels is a major focus of neural network interpretation. Although this has its merits, analyzing the results of more than a few images is tedious. In addition, users need to compare several activation channels to locate common features and determine which ones contribute towards the goal task, which is challenging for a large number of activation channels. 

\para{\challengedef{C2}. Difficulty visualizing activation similarity.}
Only visualizing channels as images side-by-side is inadequate for perceiving their similarity. 
This is due to (\rom{1}) difficulty of comparing multiple images side-by-side and (\rom{2}) ineffectiveness of human perception in comparing only the magnitudes from multiple images~\cite{jacobs_cosaliency_2010, rosenfeld_totally_2018}. 

\para{\challengedef{C3}. Difficulty identifying layers/channels contribution.}
Understanding how specific layers or activation channels contribute to the goal task is essential for debugging and optimizing neural networks. However, existing visualization techniques often lack mechanisms to quantify or highlight these contributions effectively. This challenge is further compounded when channels contribute redundantly or negatively to the task, making it difficult to isolate the most impactful components.

\subsection{Visualization Goals}
\label{sec:goals}

To address these design challenges, we have developed the following design objectives.


\para{\challengedef{G1}. Explore from abstract view to raw activations.}
To address \challenge{C1}, instead of presenting all activation channels simultaneously, we present high-level abstractions over (\rom{1}) activations per layer and (\rom{2}) the classes (for classification tasks) or pseudo-classes (for non-classification tasks). 
This will first facilitate layer-level navigation and then concentrate on interesting layers for channel-level visualization.
Abstraction of all classes will help select classes to focus more and finally navigate sample images of a single class to look for improvement opportunities.

\para{\challengedef{G2}. Compare between summaries instead of images.}
Comparing activation channel magnitudes also shows how similar they are, as the magnitudes represent the presence of features in the input.
As goal \challenge{G1} suggests, we need to show abstract information (summaries) about the activation channels. Presenting these summaries side-by-side will help users perceive the correlation and differences between the activations at the channel level, addressing \challenge{C2}. 
This approach is more effective since comparing individual values is simpler than comparing images.

\para{\challengedef{G3}. Visual analytics on patterns of activations.}
We aim to identify channels that contribute less towards goal task, addressing challenge \challenge{C3}, by guiding the user through a visual analytics process and providing information on what to expect from each visual component in the tool. 
For example, the user will start from one layer in the model, and if the activation channels in that layer look reasonable, the user will move to one neighboring layer at a time until there is a layer with unexpected behaviors (e.g., activation channels focusing on unimportant features on the input). 
As neural networks have a layer-cascading effect, retraining from that layer to output layer may fix the unexpected behavior.

\begin{figure}
    \centering
    \includegraphics[width=\linewidth]{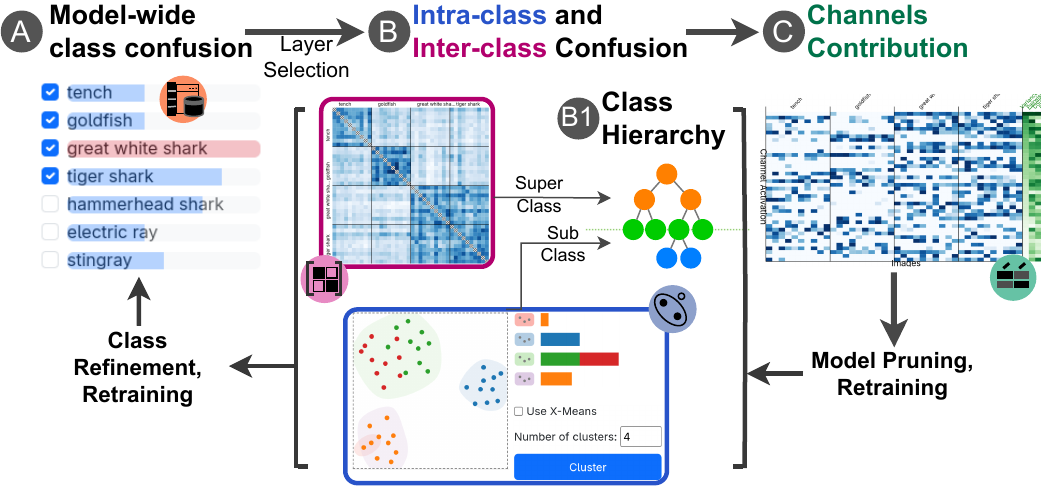}
    \caption{Overall workflow of \toolname{}. Users start the analysis with \datasetcolor{Dataset View (A)} to find classes with high confusion. Then the system shows intra-class and inter-class confusions using the two layer-level views (B), and constructs class confusion hierarchy (B1). The \heatmapcolor{Heatmap View} shows the activations of each channel (C). 
    Finally, this approach can be iteratively used to refine both the model and the dataset.}
    \label{fig:overview_flow}
\end{figure}

\para{System Workflow.} Based on the three visualization goals, the pipeline works as follows (see \Cref{fig:overview_flow}): For classification tasks, users select classes to load inputs from. For non-classification problems, users divide the inputs into groups (called \textit{pseudo-classes}) for visual separability in the tool. Upon starting the system, the model is displayed as a node-link diagram with collapsible sections in each node. The hierarchy view section automatically creates a hierarchy using super-classes and sub-classes using class confusions. Users can validate this by expanding each views of summarized outputs and compare them. For further analysis, they can view raw activations, compare inputs, and explore the layer's latent space. To track specific inputs throughout the system, clicking and focusing on any of the views will highlight the selection in all open visualizations. 



\section{\toolname{} Visualization Design}
\label{sec:viz_tool}

This section discusses the design and implementation of \toolname{}. 
We first discuss the neural network overview and summarization in \cref{sec:neural_network_overview} and \cref{sec:summarization}.
To begin, users can select classes and images in the \datasetcolor{Dataset View} (\dataset{} in \cref{fig:teaser}) and then explore the activations for those images using \scattercolor{Scatterplot View} (\scatterplot{} in \cref{fig:teaser}) and \jaccardcolor{Jaccard Similarity View} (\jaccard{} in \cref{fig:teaser}). Then, for interesting layers, the \heatmapcolor{Heatmap View} (\heatmap{} in \cref{fig:teaser}) can be used to explore activation channels. 
The implementation details are presented in supplementary materials. A demo with InceptionV3 model is publicly accessible in \href{http://155.98.19.71:9000/}{155.98.19.71:9000}.

\subsection{\texorpdfstring{Dataset View \dataset{}}{Dataset View}}
\label{sec:dataset}

For classification tasks, users can load multiple inputs from different classes. For non-classification tasks, we propose loading input groups that are expected to produce the same output from the model, which we refer to as a pseudo-class.
The \datasetcolor{Dataset View} can show model-wide activation similarity of all classes and the model's prediction for individual inputs.



\subsection{Neural Network Overview}
\label{sec:neural_network_overview}
\begin{figure}[!ht]
    \centering
    \includegraphics[width=\linewidth]{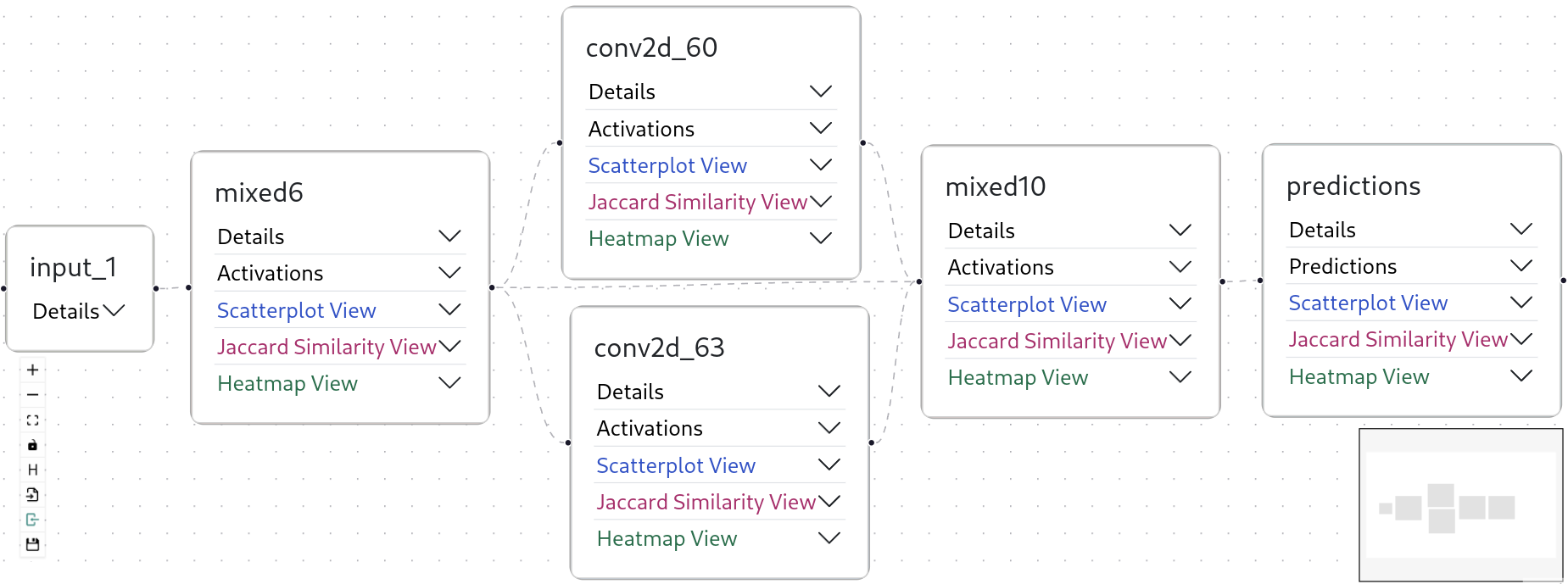}
    \caption{InceptionV3 model visualization using \toolname{}.}
    \label{fig:neuralnet-overview}
\end{figure}

The neural network overview shows the model as a graph, where each node represents a layer, and each edge represents connections between adjacent layers (see \cref{fig:neuralnet-overview}). Sugiyama layout \cite{10.1007/978-3-540-31843-9_17} is used for the initial node placement, which can be dragged to reposition. We used a detail-on-demand approach~\cite{shneiderman_eyes_1996} and added each view as a collapsible component in each node. 
To improve the visualization experience, layout navigation buttons (e.g., zoom in/out, orientation, full-screen), layout import/export buttons, minimap, and some accessibility buttons have been added. 




\subsection{Activation Channel Summarization}
\label{sec:summarization}

We create an abstraction for activation channels to address the visualization difficulty for abundant channels (\challenge{C1}).
Most neural networks' activation channels only produce high values on the presence of features in the input. This activation magnitude can be identified with only the intensity of output activation channels, regardless of the spatial location of the high-intensity pixels. Thus, we use the concept of summarization functions that quantify a channel's activation. As different non-linearity between layers (e.g., activation functions like ReLU \cite{agarap_deep_2019}, PReLU \cite{he_delving_2015}, ELU \cite{clevert_fast_2016}, SELU \cite{klambauer_self-normalizing_2017}, etc.,
and pooling \& normalization layers like max-pool, avg-pool, batch normalization, etc.) can affect the activations; we always apply summarization functions directly on the output of the layers before the non-linearity. 

Let $I \in \mathbb{R}^{w \times h}$ represent an activation channel of width, $w$, and height, $h$, where $I(x, y)$ denotes the intensity value of the pixel at coordinate, $(x,y)$. The summarization function $S(I)$ maps image, $I$, to a scalar value, $s \in \mathbb{R}$, which quantitatively represents the amount of activation in that channel. 
Though $S$ can be articulated programmatically by the user, we experimented with four instances: 
\begin{enumerate}
\item L2-norm: $S(I) = \sqrt{\sum_{x=1}^w \sum_{x=1}^h I(x,y)^2}$
\item Sum of threshold: $S(I) = \sum_{x=1}^{w} \sum_{y=1}^{h} I'(x, y)$ where, $I'(x,y) = |I(x,y)|$ if $|I(x,y)| > T$, otherwise $0$. T is determined using Otsu's algorithm~\cite{otsu}.
\item \makebox[0pt][l]{Average activation channel: $S(I) = \sum_{x=1}^w \sum_{y=1}^h |I(x,y)| / (w * h)$}
\item Maximum intensity: $S(I) = \max_{x=1}^w \max_{y=1}^h |I(x,y)|$
\end{enumerate}

\begin{figure}[h]
    \centering
    \includegraphics[width=\columnwidth]{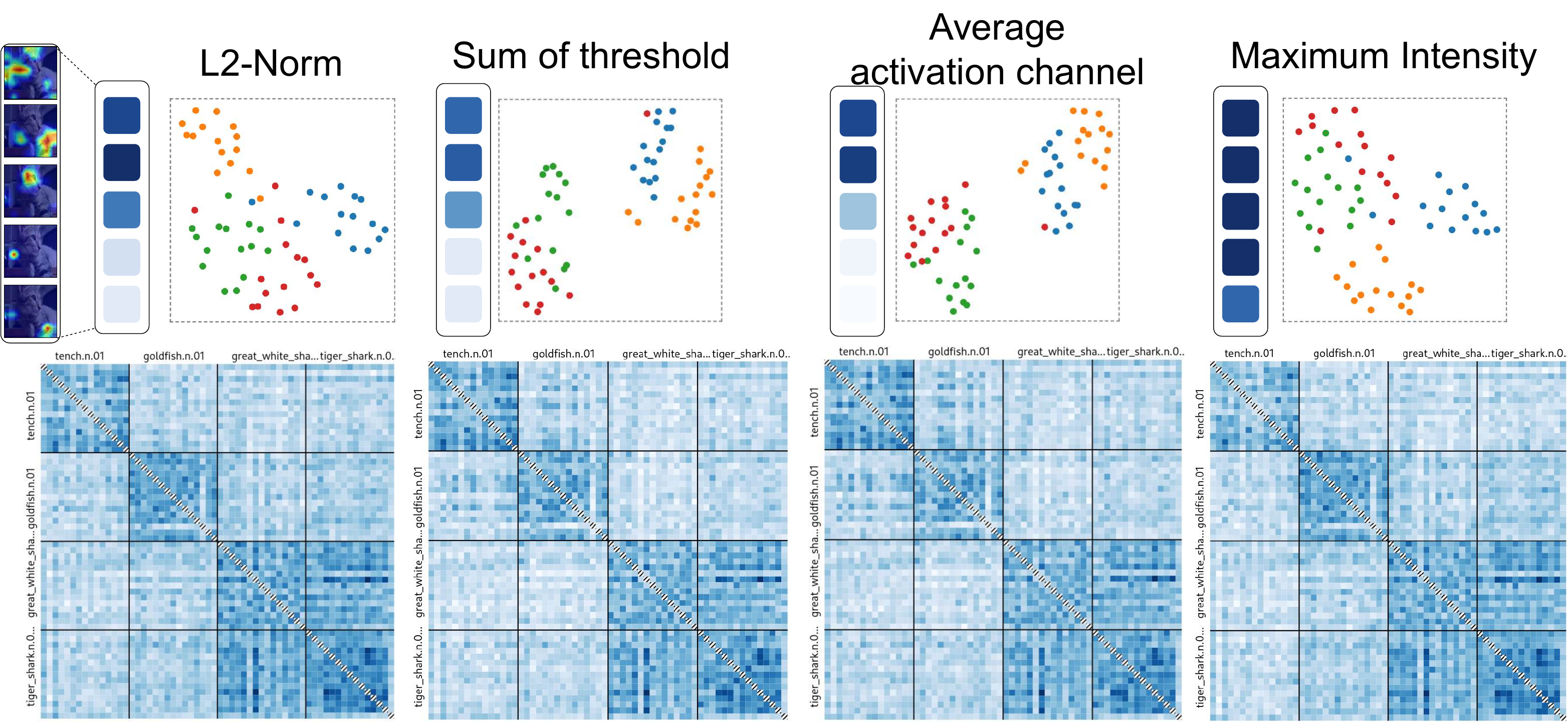}
    \caption{Effect of different summarization functions on a scatterplot, Jaccard similarity view, and activation channels. All values are globally normalized for transforming to color-scale. Geometric Threshold function }
    \label{fig:summarization-comparison}
    \vspace{-10pt}
\end{figure}

We observed that 1, 2 and 3 summarization functions work equivalently to our expectations in the \scattercolor{Scatterplot View} and \jaccardcolor{Jaccard Similarity View} (see \cref{sec:scatterplot} and \cref{sec:jaccard}). For \heatmapcolor{Heatmap View}, we observe the sum of geometric thresholding functions worked to our expectations (see \cref{sec:heatmap}). \Cref{fig:summarization-comparison} shows \scattercolor{Scatterplot View} and \jaccardcolor{Jaccard Similarity View}, and how 5 activation channels look after summarization. 
We keep the sum of threshold as the default summarization function for consistency. However, it is programmatically selectable. 

\subsubsection{Activation Distance Scatterplot View \texorpdfstring{\scatterplot{}}{}}
\label{sec:scatterplot}


The first step of goal \challenge{G1} is to navigate at the layer level. We identify layers with inter-class and intra-class confusions by visualizing the activation distances. 
This can also help reduce class confusion by identifying subclasses in the dataset (see \cref{sec:use-case-hierarchy} for a use-case).
This is accomplished by showing the result of Dimension Reduction (DR) of all activation channels of each image in a 2D scatterplot with color legends (\scatterplot{} in \cref{fig:teaser}). 
This view can help identify whether layers can group images in latent space, partially addressing goal \challenge{G3} (identifying layers' contribution).
For previous layers, points close to each other means the layer is identifying strong common features among them. 
During high-level exploration, users can expand the scatterplot view to compare multiple layers. When focusing on a specific layer, they can replace points with input images, enabling a comprehensive visualization of all inputs within a single view (see~\cref{fig:suppl-scatterplot-image} in the supplement).

\begin{figure}[!ht]
    \centering
    \includegraphics[width=0.9\columnwidth]{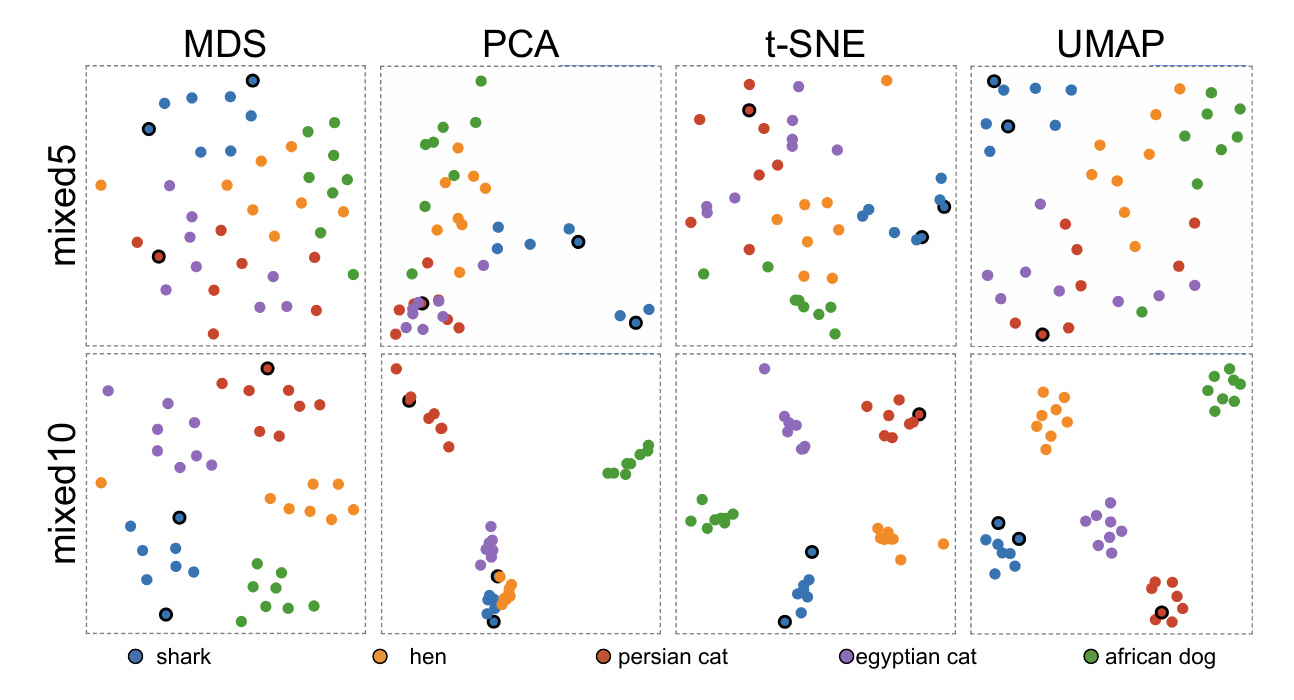}
    \caption{Four DR methods applied to two layers' summarizations of activation channels of InceptionV3 CNN model. The last CNN layer (mixed10) shows that UMAP and t-SNE preserved the global distances between classes, whereas MDS and PCA focused more on the local structure and variance.}
    \label{fig:embedding}
\end{figure}

We experimented with four DR methods for 2D embedding: PCA \cite{mackiewicz_principal_1993}, MDS \cite{kruskal_multidimensional_1964}, t-SNE \cite{maaten_visualizing_2008} and UMAP \cite{mcinnes_umap_2020}. 
MDS focuses on preserving pairwise distances, UMAP prioritizes local neighborhoods while maintaining some global structure, PCA preserves global variance, and t-SNE emphasizes local structure.
The goal of the scatterplot view is to visualize varying classes.
This addresses challenge \challenge{C2} by extracting activation distances while keeping goal \challenge{G1} by only showing an abstract overview of all inputs per layer.
We find that the clusters of t-SNE and UMAP look visually better, supported by the dimension reduction discussion in \cref{sec:related_works}. UMAP is selected as default for its faster performance and less parameter dependency (no perplexity parameter to tune).
\Cref{fig:embedding} shows results with the best settings per DR method. As initialization of these DR methods greatly changes the result, we provide ways to refresh with new initialization so users can quickly see multiple results with different initializations.

\begin{figure}[!t]
    \centering
    \includegraphics[width=\columnwidth]{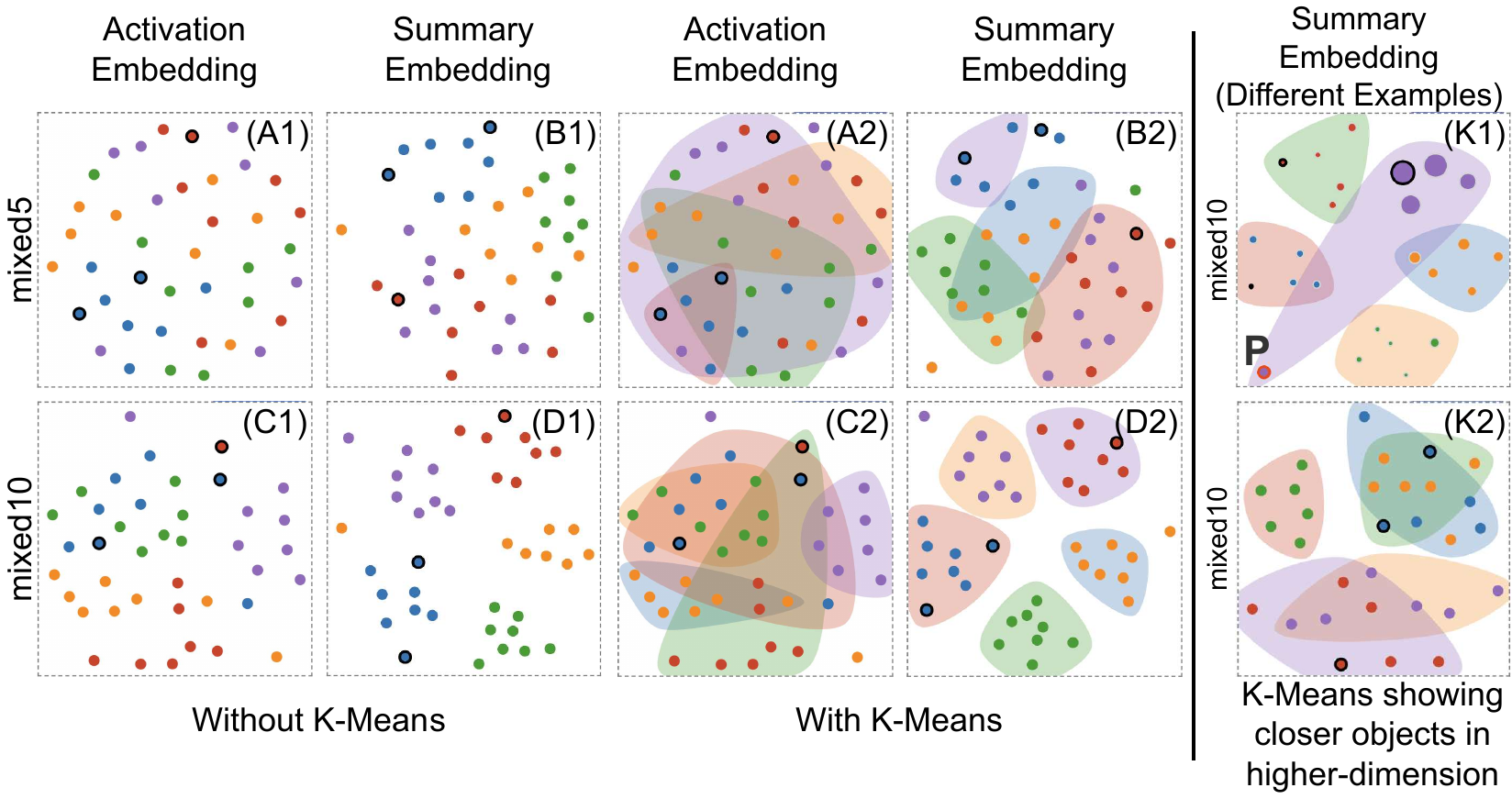}
    \caption{(A1-D2) Comparison between \scattercolor{Scatterplot Views} using whole activation channel vs. summarization of activation channel. 
    (K1) The purple point with a red border, \textbf{P}, is far from its cluster (purple points cluster), but K-Means clustering shows that they are close.
    (K2) Purple and Red points' clusters are ambiguous in 2D, but K-means show they are separable in higher dimensions.
    }
    \label{fig:summery_full_embedding}
    \vspace{-10pt}
\end{figure}

Dimensionality reduction (DR) methods alone cannot fully address challenge \challenge{C2}, as they often distort distances between activations. Points close in high-dimensional space may appear distant after projection, resulting in misleading interpretations. 
Even in higher dimensions, nearest-neighbor algorithms are meaningful when they form clusters~\cite{nn_meaningful}. 
To mitigate the structural distortion caused by DR in visualization, we incorporate X-means clustering using Euclidean distance~\cite{xmeans}. While we also tested DBSCAN, its reliance on point density made it less effective and interpretable in high-dimensional spaces. Finally, the interface uses K-means when users provide a value of $K$. While the number of clusters in X-means often aligns with the number of (pseudo-)classes, 
allowing users to set $K$ helps analyze class separability by checking if clusters align with input class labels.
We draw these clusters with hull enclosures using Catmull-Rom curves \cite{barnhill_class_1974}. 
\Cref{fig:summery_full_embedding}(K1) shows that even though the point, \textbf{P}, is far away from its class members (purple points in the purple cluster), the enclosure shows they are still close enough to form a cluster in high dimension. Moreover, \cref{fig:summery_full_embedding}(K2) shows overlapping clusters; however, X-Means enclosures show they are separable in high dimensions. 

We add a horizontal barchart showing the distribution of classes within each cluster. 
A cluster containing well-distributed samples from many classes may indicate two cases: high inter-class confusion or poor centroid initialization. Users can re-run clustering with different seeds through the interface to distinguish between these cases.

We experimented with the application of DR in two ways. First, we flattened the activation channel from a 2D scalar field, $C_{ij}^l \in \mathbb{R}^{w \times h}$, into a vector $V_{ij}^l \in \mathbb{R}^{wh}$, where $C_{ij}^l$ is the $i$th channel (of size $w \times h$) for $j$th input image at layer $l$, and applied DR using the Euclidean distances between the concatenation of all vectors of each image. Second, we use the summary of activation channels (i.e. $S(C_{ij}^l) \in \mathbb{R}$) and use the distance between the vector of summaries of all channels of each image. \Cref{fig:summery_full_embedding} shows results of DR with and without using summarization functions. 
We observe that L2-norm has the best result for all DR methods and show all the scatterplots with that summarization function. 

\subsubsection{Activation Jaccard Similarity View \texorpdfstring{\jaccard{}}{}}
\label{sec:jaccard}

While reducing dimensions can effectively create abstract visualizations for layers (\challenge{G1}), DR and cluster enclosures do not show the original distances between activations. Hence, they are not enough to completely satisfy \challenge{G2}.  
We incorporate a similarity matrix that visualizes the exact distance between activations of a layer.


Similar to \scattercolor{Scatterplot View}, we tested measuring the distances between activation channels of each image, both with and without summarization functions. Without summarization, we compute the pixel-wise sum of distances between corresponding activation channels for each image. This results in intra-class distances (diagonal of the similarity matrix) being nearly identical to inter-class distances, offering limited insights. In contrast, applying summarization functions produces clear differences between diagonal and non-diagonal cells in the similarity matrix. Therefore, we consistently use summarization functions for the similarity matrix.

\begin{figure}[!ht]
    \includegraphics[width=\linewidth]{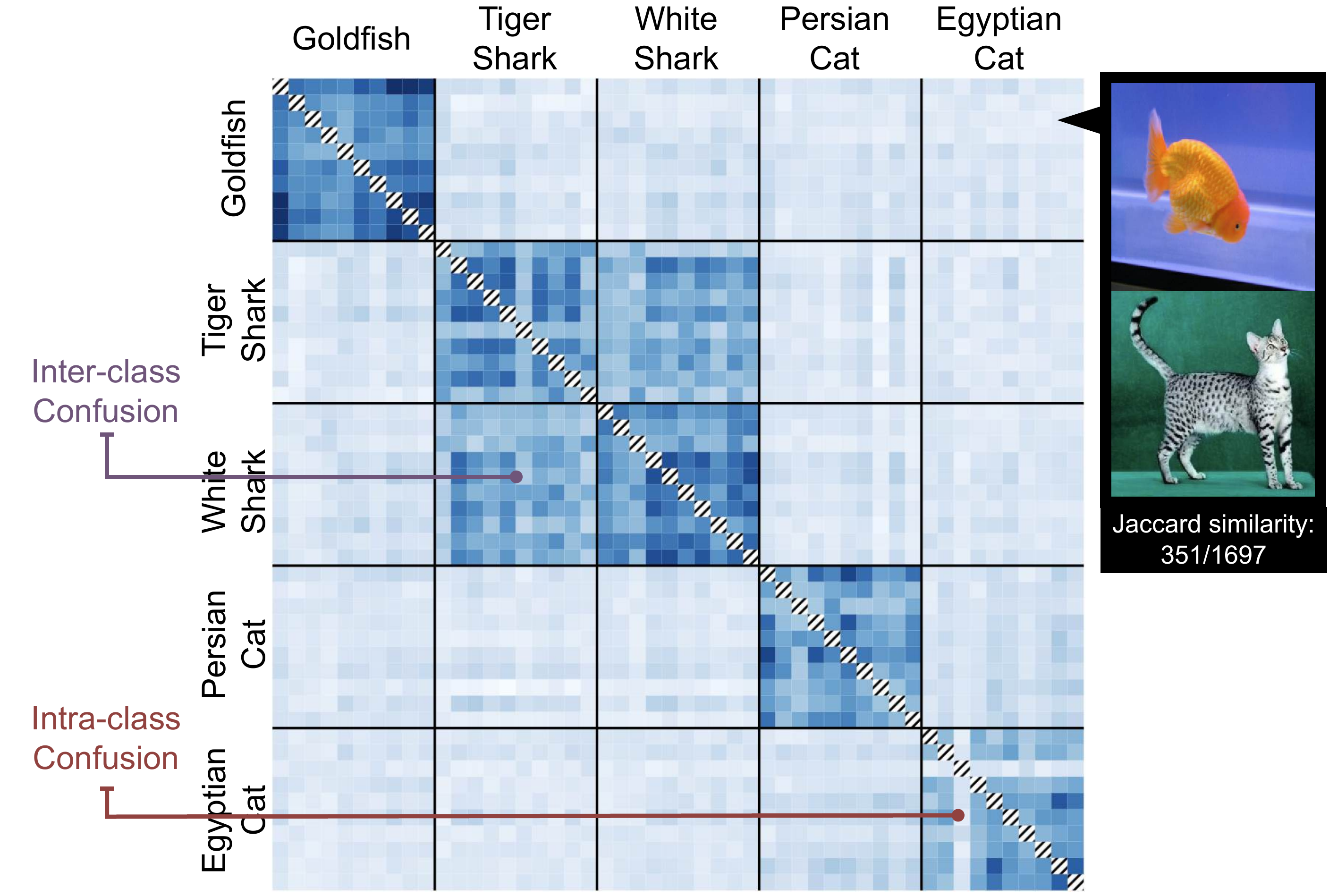}
    \caption{Demonstration of \jaccardcolor{Jaccard Similarity View}. Darker colors in the distance between two classes (non-diagonal cell groups) represent inter-class confusion and lighter colors in diagonal cell groups represent intra-class confusion.}
    \label{fig:jaccard}
    \vspace{-10pt}
\end{figure}

We first apply summarization functions to each channel and create vectors, $V_{ij}^l$ (as in \cref{sec:scatterplot}), for each $i$th image. 
Then, we pick the top $A_\eta$ activation channels from all $V_{ij}^l$ to form a set where $A_\eta = \lceil \eta k \rceil$, $0 < \eta \le 1$, and $k$ is the number of activation channels per image for that layer. Here, $\eta$ is a tuning parameter: higher values of $\eta$ pick more channels, including those that loosely identify features, and lower values of $\eta$ pick fewer activation channels, those that highly activate for certain features. 
Let the set of top $A_\eta$ activation channels for $i$th image at layer $l$ be $S^l_i$. We consider the channels to be "fully activated" only if it is a member of $S^l_i$. 
The Jaccard similarity coefficient \cite{jaccard1902distribution} matrix at layer $l$, $J^l$, is defined by

\begin{equation}
J_{ij}^l =
    \underbrace{|S_i^l \cap S_j^l|}_{\substack{\text{\# of channels} \\ \text{activated for \textbf{both} images}}} / \underbrace{|S_i^l \cup S_j^l|}_{\substack{\text{\# of channels} \\ \text{activated for \textbf{either} image}}}
\end{equation}
To address challenge \challenge{C2}, we visualize this Jaccard similarity matrix (\jaccard{} in \cref{fig:teaser}) as a heatmap. 
The color scale is optionally normalized in the range of 1st and 99th percentile of activation channel summaries to reduce sporadic bright rows and columns in the matrix caused by erratic input data.
The rows and columns of the same classes are grouped using separator lines.
Clicking on any cell in the matrix will show the two inputs and the number of common channels activated (see \cref{fig:jaccard}).

The \jaccardcolor{Jaccard Similarity View} addresses goal \challenge{G2} by pointing intra-class and inter-class confusions within a layer. 
If two classes share many activation channels, their similarity positions (non-diagonal groups) will have darker colors, representing high inter-class confusion. Oppositely, lighter colors in diagonal groups represent intra-class confusion.
\Cref{fig:jaccard} shows that class \textit{Egyptian Cat} has high intra-class confusions and class \textit{Tiger Shark} \& \textit{White Shark} has high inter-class confusions for this layer. 

\para{Automatic Confusion Hierarchy Generation.}
Using \scattercolor{Scatterplot View} and \jaccardcolor{Jaccard Similarity View}, the system automatically generates and visualizes class confusion hierarchy for each layer. The super-classes are created with high inter-class confusion values, and sub-classes are created by analyzing class distribution in multiple clusters (see "Confusion Hierarchy" section in layer \texttt{fc1} in~\cref{fig:teaser}).

\subsubsection{Activation Heatmap View \texorpdfstring{\heatmap{}}{}}
\label{sec:heatmap}

After navigating at the layer level, goal \challenge{G1} suggests exploring raw activation channels. We identify interesting layers using the \scattercolor{Scatterplot} and \jaccardcolor{Jaccard Similarity} views and then use the \heatmapcolor{Activation Heatmap View} (\heatmap{} in \cref{fig:teaser}) to visualize activation patterns at the granularity of channels. In this \heatmapcolor{Heatmap View}, each column represents an input image, and each row represents an activation channel. 
As goal \challenge{G2} suggests, comparing only the summaries of activations is easier than comparing thousands of images.
Furthermore, as the \scattercolor{Scatterplot View} addresses one part of goal \challenge{G3} (identify layers' contribution), the \heatmapcolor{Heatmap View} addresses the other part, identifying activation channels' contributions by comparing the activation pattern at each layer from inputs of different classes.

\begin{figure*}[!ht]
    \centering
    \includegraphics[width=0.9\linewidth]{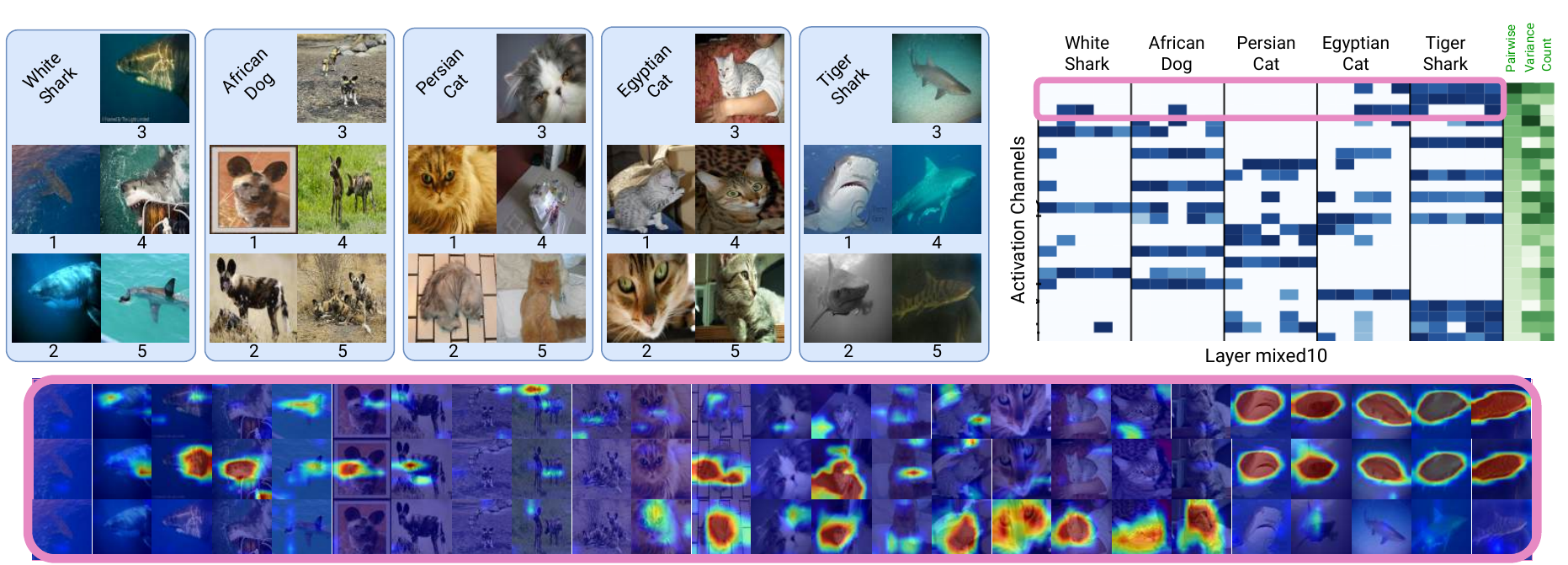}
    \caption{\heatmapcolor{Heatmap View} of layer mixed10 in InceptionV3. The top 3 activation channels are shown as overlays at the bottom. The heatmap's stripes (dark color in all columns of a class in a row) show that activation channels at those rows detect class-identifying features.}
    \vspace{-10pt}
    \label{fig:heatmap}
\end{figure*}

Images (columns) are grouped by the provided classes. To reduce the dominance of low activations, activation summary values are thresholded at the 10th percentile before applying the colorscale. Clicking a cell displays the activation channel overlaid on the input image, calculated as follows.

We assume the maximum and minimum pixel value of all activation channels at layer $l$ is $G_{max}^l = \max_{\forall i,j,x,y} C_{ij}^l(x,y)$ and $G_{min}^l = \min_{\forall i,j,x,y} C_{ij}^l(x,y)$ respectively, where $C_{ij}^l \in \mathbb{R}^{w \times h}$ is the $i$th channel of $j$th image. We normalize the activation channel, $\tilde{C}_{ij}^l(x,y)$, by scaling between $G_{min}^l$ and $G_{max}^l$ and then apply bilinear interpolation to resize to the input image. This is then converted into RGB colors using a color mapping function $M_c$, where $c \in \{R,G,B\}$. Finally, we blend this RGB activation with the input image channels, $f_c$, to get the final output, $O_c$. We use $\alpha$ as a blending factor. For default, we take the Jet diverging colormap and $\alpha = 0.6$.

\begin{equation}
O_c(x,y) = \alpha M_c(\tilde{C}_{ij}^l(x,y)) + (1 - \alpha) f_c(x,y)
\end{equation}

In \cref{fig:heatmap}, we observe the highest number of consistent stripes (dark blue for all columns in a class group) when using the sum of threshold as summarization function for a high confidence InceptionV3 model. Thus, we use this as default summarization function.

Showing thousands of rows representing activation channels can obfuscate the visualization without showing any summary or aggregated metric. Through various experiments, we add 3 additional metrics for ordering these activation channels: (\rom{1}) variance, (\rom{2}) sum of class-pairwise distance, and (\rom{3}) edge weights from the previous layer. 
Let $\mu_{i}^l$ be the average of all $S(C_{ij}^l)$, summary of $i$th activation channels (row $i$) of all $n$ images (all $j$ columns) at layer $l$. 
The variance is ${\sigma_i^l}^2 = \frac{1}{n} \sum_{j=1}^n (S(C_{ij}^l) - \mu_i^l)^2$.
We assume the number of classes as $\mathcal{C}$, and $\mathcal{C}_{ui}^l$ represent the vector of summary of $i$th channels of all images in class $u$. We compute $d(\mathcal{C}_{ui}^l, \mathcal{C}_{vi}^l)$ to be the Euclidean distance between channel summaries of two classes. Then, the sum of class-pairwise distance is calculated by

\begin{equation}
\zeta_i^l = \sum_{j_1=1}^{\mathcal{C}-1} \sum_{j_2=j_1+1}^\mathcal{C} d(\mathcal{C}_{ij_1}^l,\mathcal{C}_{ij_2}^l)
\end{equation}

We calculate the L2-norms for each filter as the edge weights. These aggregated metrics are added as the last columns to the heatmaps (green columns in \cref{fig:heatmap}). The sum of class-pairwise distance is used as the default ordering of channels, as it shows the highest number of stripes at the last layer for good models.

For class-separating channels, each column group should have exclusive stripes. For example, layer mixed10 of \cref{fig:heatmap} has 2 stripes in the top two rows for \textit{Tiger Shark} class, meaning they are looking for \textit{Tiger Shark} specific features. 
When an activation channel is activated sporadically for all images of all selected classes (no stripe patterns), it does not contribute to the final task. We use this behavior to achieve goal \challenge{G3} (identify channel contribution) using this expectation of similar activations for each feature and find the channel's contributions in a use case of section \cref{sec:use-case-channel-contribution}. 
\section{Implementation Details}
\label{sec:implementation}



\para{Open-source and Reproducible Implementation.}
\toolname{} is an open-source web-based application deployable on any local machine. 
The system consists of a decoupled backend and a frontend. The backend, built with Python (FastAPI, TensorFlow, PyTorch), runs the model and returns results. The frontend, built with TypeScript, uses React, Redux, and D3 for visualization. API type safety is enforced using Beartype. 
Most computations for global values (e.g., maximum and minimum summary values per layer, average class activation distance in \datasetcolor{Dataset View}) are calculated for the whole dataset, as these can be inaccurate if calculated for only user selection.

\para{Scalability of \toolname{}}
We tested the system with an RTX 2060 GPU and without the GPU (i.e., CPU only). Both the complexity of backend and frontend is linear to the number of inputs. We loaded up to 700 images with 10 classes in ConvNeXt-XL~\cite{liu2022convnet} model with $\sim$350M parameters and the system was usable with minor slowness. Loading more images typically results in ineffective visualization.

\para{Supported Neural Network and Dataset Types.}
\toolname{} supports any image-based model implemented in or converted to Tensorflow, PyTorch or ONNX format models.  
Any categorical dataset (having class or pseudo-class tagged for each input) including \texttt{tensorflow\_dataset} and \texttt{torchvision.datasets} are supported.

\para{API documentation and Public Access.}
The API documentation is provided in the link \href{https://channelexplorer.netlify.app/}{channelexplorer.netlify.app}. The project's source code can be found in \href{https://anonymous.4open.science/r/ChannelExplorer}{anonymous.4open.science/r/ChannelExplorer}. Dockerized containers are provided for easier reproducibility. A demo with the ImageNet dataset and InceptionV3 model is publicly accessible in \href{http://155.98.19.71:8000/}{155.98.19.71:8000}. The project is pushed to PyPI~\cite{pypi} and can be installed by "\texttt{pip install channelexplorer}". 
\section{Use Cases of \toolname{}}

We discuss four use case scenarios. 
\subsection{Scenario 1: Enhancing Class Hierarchy}
\label{sec:use-case-hierarchy}

Deep neural networks often capture more information than needed for the task~\cite{AyindePruning2018}, which may be used to find inherent groupings of the dataset. To demonstrate this, we applied the visualization technique to a CNN classifier and a super-resolution GAN model.

\para{Class Hierarchy in CNN.}
We use the proposed visualization to create a hierarchy of 5 classes of the ImageNet dataset and InceptionV3 model's mixed10 layer. Four related classes (\textit{Ibex}, \textit{Bighorn}, \textit{Bison}, and \textit{Ox}) are selected for the hierarchy, and one class (\textit{cat}) is selected to create baseline distance in the \jaccardcolor{Jaccard Similarity View}. 

\begin{figure}[!b]
    \centering
    \includegraphics[width=0.95\columnwidth]{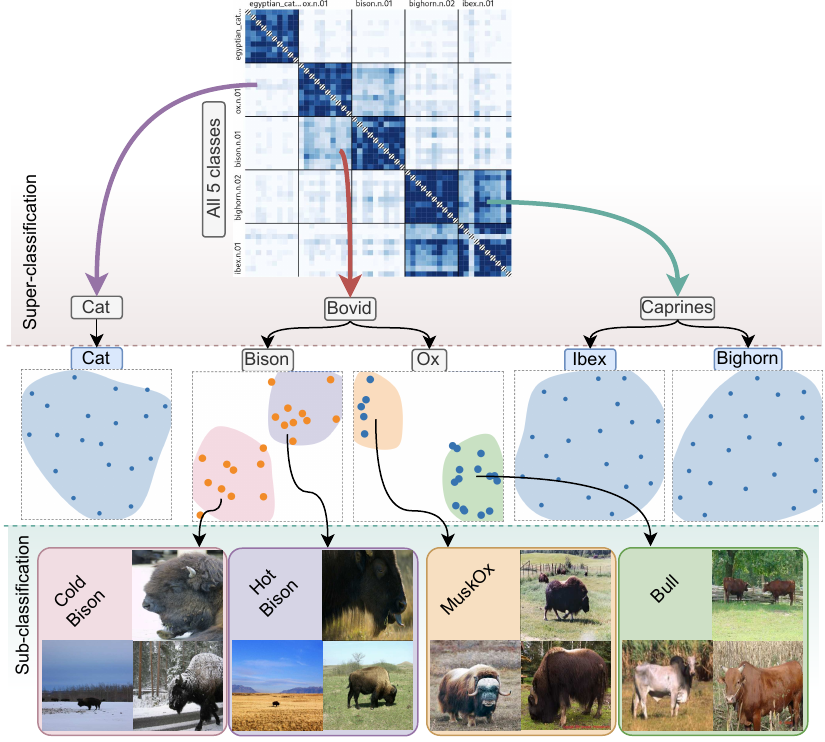}
    \caption{Starting from 5 classes, we create 3 super-classes (Cat, Bovid, and Caprines). 
    In the \textit{Bison}'s \scattercolor{Scatterplot View}, two clusters show examples of \textit{Bison}s in Cold \& Hot weather. 
    Similarly, in the \textit{Ox} \scattercolor{Scatterplot View}, one cluster represents furry, barrel-shaped bodies (MuskOx breed), and the other represents less furry, cow-like bodies (Bull), leading to the creation of two subclasses.}
    \label{fig:class-hierarchy}
\end{figure}

\para{Super-classification in ImageNet.}
The \jaccardcolor{Jaccard Similarity View} in \cref{fig:class-hierarchy} shows a high similarity between \textit{Bison} \& \textit{Ox} and \textit{Ibex} \& \textit{Bighorn} among 10 images per class. Hence, we merged them into two super-classes, naming \textit{Bovid} and \textit{Caprines}, respectively.

\para{Sub-classification in ImageNet.}
Upon loading 20 images, the scatterplots in \cref{fig:class-hierarchy} show no clusters in \textit{Cat}, \textit{Ibex}, and \textit{Bighorn} classes, but strong bi-clusters in \textit{Bison} and \textit{Ox} classes. 
The images from \textit{Ox} class show (by clicking on points) that one cluster contains Muskox images, and another has Bull. The cluster distance indicates that the model has to determine two different \textit{Ox} specifications to classify both breeds of Oxen, which is intra-class ambiguity. So, we divide it into two subclasses named Muskox and Bull. Similarly, class \textit{Bison} is divided into two subclasses, \textit{Bison}s in Cold and Hot environments. More sub-classification instances are presented in supplementary materials.

\para{Refining Model Classification.}
We used the model inference on the entire dataset to create two new subclasses: Bison in hot \& cold environments and Bull \& Muskox. We then manually verified whether these subclasses semantically aligned with their respective classes, making necessary corrections. Finally, we retrained the model’s last layer with 1002 classes. The updated model maintained the same accuracy and speed while expanding its classification capacity from 1000 to 1002 classes. This demonstrates that the visualization approach, combined with minimal human intervention, can increase the model’s classification ability. Further details on the transfer learning process are explained in the supplementary materials.

\para{Class Relationships in GAN.}
Similar to InceptionV3, we load images from classes \textit{parachute}, \textit{coral reef}, \textit{cliff}, \textit{orangutan}, and \textit{persian cat} of ImageNet dataset into SRResNet~\cite{srresnet2017}, a Generative Adversarial Network for super-resolution imaging. Although the generator is used for inference, we show what the discriminator model \emph{sees} and whether inter-class relationships exist during the identification of whether an input image is from a distribution of high-resolution or synthetic images.

The \jaccardcolor{Jaccard Similarity View} of 10 images per pseudo-class from the final convolutional block shows high similarities between \textit{coral reef} \& \textit{cliff} and \textit{orangutan} \& \textit{persian cat}. While these similarities may be surprising due to the lack of semantic relationships between them, we sensibly assume they likely reflect the discriminator's reliance on texture-based similarity. 
Furry textures in \textit{orangutan} \& \textit{persian cat} and stone-like textures in \textit{coral reef} \& \textit{cliff} may dominate during discriminator training, as the generator might imperfectly upscale them.
The \textit{parachute} class, mostly dominated by clear sky backgrounds, is added to standardize the colorscale of \jaccardcolor{similarity matrix} to improve visual clarity (similar to the use of \textit{cat} class in InceptionV3). Supplementary materials provide corresponding \jaccardcolor{Jaccard Similarity View}.

Our visualization technique, applied with SRGAN, highlights how model inputs can show hierarchical relationships based on feature-level patterns, even if they lack semantic significance.

\subsection{Scenario 2: Identification of Mislabeled Images}
\label{sec:use-case-mislabeled}
\begin{figure}[ht]
    \centering
    \includegraphics[width=\columnwidth]{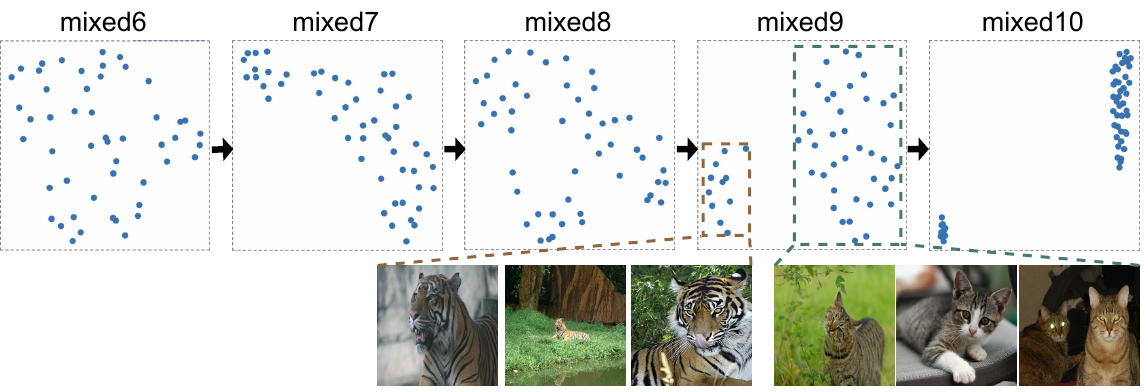}
    \caption{Taking 50 images of class \textit{Tiger Cat}, layers mixed8 to mixed10 have consistent cluster formation. One of those clusters is images of class \textit{Tiger}, which is mislabeled.}
    \label{fig:tigercat-mislabel}
\end{figure}

While browsing through the activations, we find that the sub-classification (or outlier) may also happen due to wrong labeling of images. 
For example, the \scattercolor{Scatterplot View} shows two sub-classes for the class \textit{Tiger Cat} (class index 282) in the ImageNet dataset. 
However, one of the clusters contains \textit{tigers} (class index 292), which is mislabeled as \textit{Tiger Cat} (see \cref{fig:tigercat-mislabel}). 

This separation behavior occurs in layers that identify features present in one input group (cat) but absent in another (tiger). As shown in~\cref{fig:tigercat-mislabel}, this distinction becomes noticable after layer mixed9, where activation channels selectively activate for one group while deactivating for the other. The \heatmapcolor{Heatmap View} further illustrates this phenomenon through distinct stripe distributions between the two groups.


\subsection{Scenario 3: Identify Activation Channels' Contribution}
\label{sec:use-case-channel-contribution}
\begin{figure}[ht]
    \centering
    \includegraphics[width=0.9\linewidth]{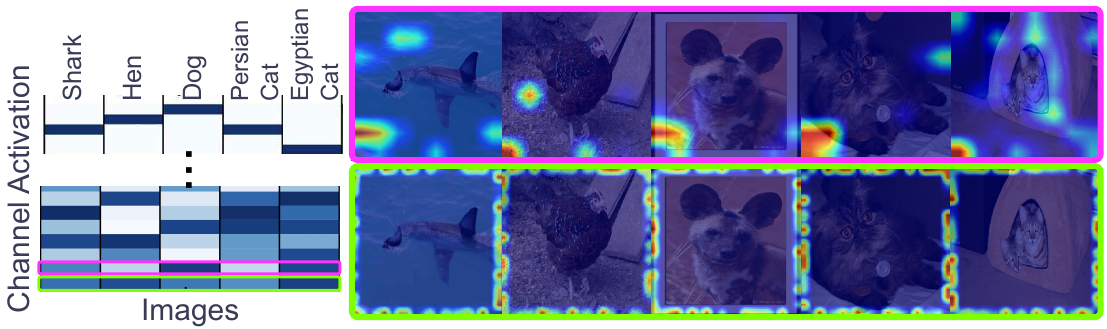}
    \caption{Activation channels at the bottom of \heatmapcolor{Heatmap View} that identify unimportant features for the classification. The first row always activates in the bottom left corner. The second row activates at the border regardless of the input image.}
    \label{fig:nonsense_channels}
    \vspace{-10pt}
\end{figure}

A channel is activated when it finds the presence of a feature. The input might have some features that do not contribute towards the goal task (i.e. background of an object for object detection task), rendering the channels that detect those features less important. 
From the \heatmapcolor{Heatmap View} with the sum of class-pairwise distance order, we observe sporadic activations of channels throughout all the classes at the bottom of the heatmap for the mixed10 layer of InceptionV3. These channels mostly focus on the wrong regions of inputs, and the sporadic activation means they are not detecting class-identifying features.
\cref{fig:nonsense_channels} shows overlays of the last 2 rows of channels focusing around the border region.
Removing these channels may improve model's memory usage and/or inference speed without compromising accuracy. 
We evaluated this by incrementally removing channels from layers and calculating the mAP over the ImageNet eval dataset. We were able to remove the bottom 74\% of the channels without affecting mAP, yielding 12.1\% interference time improvement in CPU. For GPU, inference time remained the same until removing 70\% channels, after which it improved by 50\%. The memory usage was linear to the number of channels removed for both CPU and GPU. 
We compared our interpretable manual pruning with two variants of VGG16 pruned using established techniques: (1) unstructured magnitude-based weight pruning, and (2) structured filter pruning based on norm values, both reducing 40\% of weights in the layers. We found that both pruned models have accuracy comparable to our manually pruned model. However, our method offers the additional advantage of selectively fine-tuning the accuracy trade-offs between classes.
Other results of the removal of different channels and details of pruning method comparison are shown in supplementary materials.  

\subsection{Scenario 4: Locating Latent States' Position in SD}
\label{sec:use-case-stable-diffusion}
\begin{figure}[ht]
    \centering
    \includegraphics[width=\linewidth]{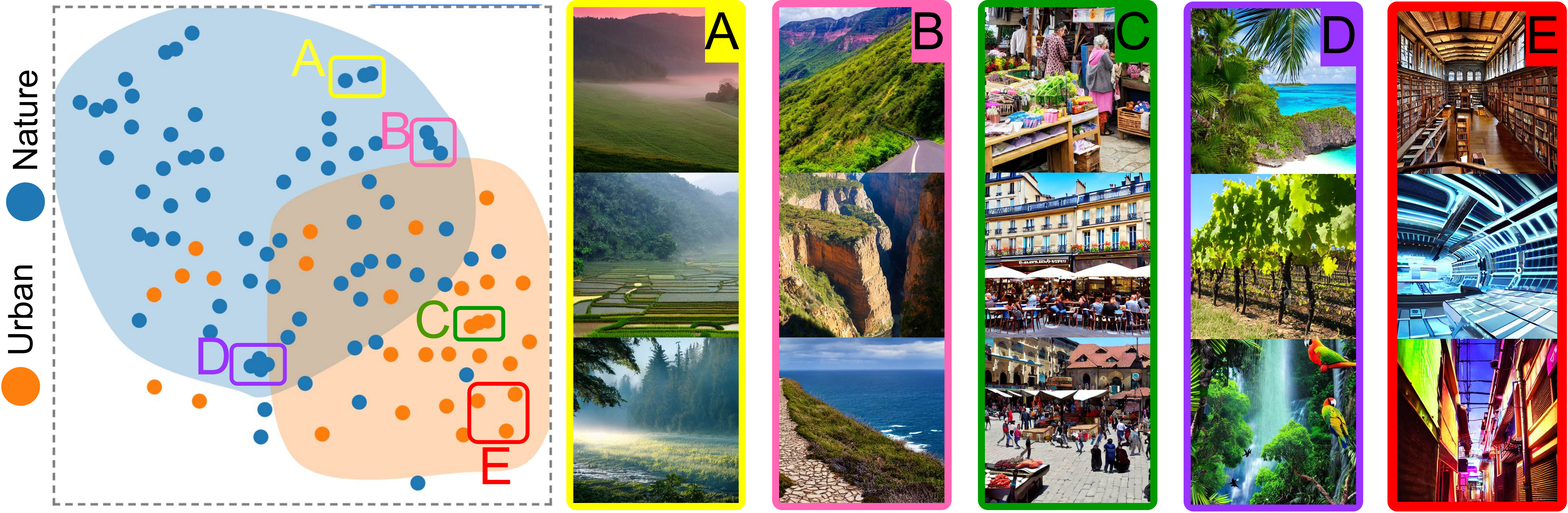}
    \caption{\scattercolor{Scatterplot View} of at 49th timestep (out of 50) output of mid\_block layer of the stable diffusion model for 100 scene descriptions. 5 groups of points are shown on the right. Column A has misty open space as a common feature. Column B has cliffy mountains, C has dense crowds, and D has large green trees as common features. No relations were found between images in E, except they are of an urban region. All input texts are added to supplementary materials.}
    \label{fig:scene-stable-diffusion}
    \vspace{-10pt}
\end{figure}

Understanding the latent position of the output of a model is crucial to determining the quality of the output. To determine the latent position of input in the Stable Diffusion (SD) model, we generated 100 text descriptions of different scenic places using LLM and focused on the output of the 'mid\_block' layer of the U-Net model.
To facilitate the visualization, we manually assigned pseudo-classes, \textit{Nature} and \textit{Urban}, to the descriptions. The \scattercolor{Scatterplot View} shows a clear separation between these two classes. 
Inspecting closely grouped images, we observed strong conceptual similarities (see \cref{fig:scene-stable-diffusion}). 
If a newly generated image lies outside these clusters, we can assume that its features are different from those represented by the 100 prompts.
This demonstrates that, with sufficient samples, \toolname{} can locate a newly generated image's position in the latent space distribution of related prompts. However, we also sometimes observed instances of visually close points lacked meaningful semantic similarity, highlighting limitations of projection-based views.

\subsection{User Evaluation of \toolname}
\label{sec:evaluation}

Our evaluation sessions lasted one hour, and the participants were nine graduate students (P0-P8) who took a graduate Deep Learning course. P0-P3 are working with Deep Learning, P4-P5 are studying Computer Science, and P6-P8 are from other fields. 
The sessions took place in person for P0, P2, P4, and P6 and virtually for others.
One author functioned as the facilitator, leading each session.

The evaluation is done with a reduced layer visualization of the InceptionV3 model and ImageNet dataset. 
Five images of five classes from \cref{fig:heatmap} are used: \textit{Great White Shark}, \textit{Tiger Shark}, \textit{Persian Cat}, \textit{Egyptian Cat} and \textit{African Hunting Dog}. 
Four tasks were performed during the evaluation. More thorough documentation of the procedure can be found in the supplement.


\para{T1: Load the dataset and guess four potentially confusing images.}
Participants were asked to load the 25 images from \cref{fig:heatmap} using the \datasetcolor{Dataset View}. 
Without knowing model prediction, they guessed 4 images that might be confusing.
This familiarized participants with the tool and later justified their assumptions. They were also asked some questions (input/output shapes, number of channels etc.) to confirm their knowledge of CNN and familiarity to the tool.
All except P5 and P7 found the 1st \textit{White Shark} and 4th \textit{Persian Cat} images potentially confusable with the 3rd \textit{Tiger Shark} and 3rd \textit{Egyptian Cat}, respectively. However, P5 and P7 noted confusion between the 1st \textit{Persian Cat} and 2nd \textit{Egyptian Cat} images.

\para{T2: Identify misleading layers for selected classes using Scatterplot View.}
Following an explanation of \scattercolor{Scatterplot View}, participants were asked to select confusing images from the scatterplot.
This task evaluates whether the variation of \scattercolor{Scatterplot View} using different initialization seeds or the perception of clusters affects users' decisions~\cite{10308641}.
P1-2 and P4-8 pointed to the 1st image of \textit{White Shark} whereas 
P0 and P3 selected the 5th \textit{Tiger Shark} and 5th \textit{Persian Cat}, respectively.
Then, they were asked to track their selections one previous layer at a time and identify a layer where the confusion becomes less noticeable. They hypothesized that 
retraining all those observed layers may improve classification accuracy.
All participants selected layers from mixed6 to mixed10 (except mixed8); however, their responses widely varied. 


\para{T3: Identify class confusion using Jaccard Similarity View.} 
After explaining the \jaccardcolor{Jaccard Similarity View}, all participants agreed that class \textit{African Dog} has the least confusion. P0, P1, P3, P5, and P7 indicated that the \textit{Persian Cat} has the highest intra-class confusion. Remaining participants observed \textit{White Shark} has the highest intra-class confusion. All participants agreed that \textit{White Shark} and \textit{Tiger Shark} have the most inter-class confusion.

\para{T4: Identify misleading activation channels using Heatmap View.}
The participants were explained the activation channels' contribution by the facilitator and were asked to identify channels that activate wrongly for the classification from the \heatmapcolor{Heatmap View}. 
All participants responded with activation channels that are at the bottom half of the view. 
We removed all selected activation channels, and the model performance was unchanged.

\para{Findings.}
The evaluation process shows that the variations of DR methods, KMeans clustering, and cluster perception do not affect the decision to identify the target layers or channels. 
Additionally, \toolname{} requires a deep understanding of neural networks for effective use.
Finally, the evaluation reveals that \toolname{} cannot exactly locate a faulty layer or activation channel; rather, it provides a range of possible regions in the model.

\section{Limitations of \toolname{}}
\label{sec:limitations}

\toolname{} overcomes the challenges in \cref{sec:challenges}, however it has some of its own limitations. 
While the backend is scalable as its complexity grows linearly with the number of inputs, the visualizations become less effective beyond about 100 inputs per class. At that point, the scatterplot becomes overcrowded, and the heatmap and Jaccard similarity views have too many columns, making visual comparison difficult.
The \heatmapcolor{Heatmap View} identifies less contributing activation channels for selected classes, but some of those channels might still be useful for other classes. 
Defining pseudo-classes to analyze earlier layers in the network is also challenging, as these layers capture basic features (shapes, textures, etc.) that are not task-related. 
While in \cref{sec:use-case-hierarchy}, we show merging classes can improve model performance, some tasks may not allow changing the number of classes.
Although any model can be loaded and visualized, all the views work best on image-based layers (e.g., CNNs, transpose convolutions, residual blocks, pooling layers) due to our approach of using summarization functions. This limits applicability to other layer types like embedding layers, transformers, or attention layers. In advanced models like Stable Diffusion, which operate iteratively, the current system can only visualize a single user-defined iteration at a time. Comparing the same layer across iterations requires launching multiple tool instances.
Finally, class-separability still remains a difficult and open problem. Like other state-of-the-art tools, \toolname{} cannot fully automate problem diagnosis or apply fixes. Some level of human interpretation and intuition is still essential. 

\section{Discussion and Future Work}
\label{sec:future_work}

\para{Real-time Model Mutation.}
\toolname{} currently provides insights into the activations without any changes to the model. 
Enhancing the tool to allow simultaneous channel addition or removal during visualization would significantly improve usability.
We are working on adding capabilities to \emph{disable} certain activations channels in different layers and see how the model behaves.

\para{Automatic Class Confusion Avoidance.}
While \toolname{} already has class confusion hierarchy tree generation pipeline, its ability to determine activation channels contribution still requires manual inspection of each cell (to see if a channel is focusing on a wrong location at input). 
We aim to apply automatic pattern recognition and apply vision models to identify the semantics of each activation channel in \heatmapcolor{Heatmap View}. This will drastically reduce the human-in-the-loop process to inspect what the channel is looking for.

\para{Insights from Activation Vectors.}
Activation Vectors can provide different insights than activation channels (see \cref{sec:background} and \cref{fig:basic-advance-features}). We want to explore the visualization opportunities with activation vectors to further enrich model optimization possibilities.

\para{Support beyond Image-based Layers.}
It has been shown that transformer and attention-based models like Vision Transformers (ViT) outperform models with image-based layers in many computer vision tasks~\cite{vitmodel}. 
For example, ViT models first divide the input image into patches and then use an attention mechanism on the sequence of embeddings. 
While it is possible to visualize the first layer of patch embeddings without any modification to the tool, this does not yield meaningful insights, as these embeddings exist in a high-dimensional latent space rather than directly
corresponding to spatial feature activations. Extending our research to visualizing attention-based representations can help cover more neural networks in the future.

\section{Conclusion}
Image-based layers play a critical role in many neural network architectures for decision-making tasks, making their proper training and inference with the right dataset essential. Image-based layers are also more interpretable and can be mapped on the input more easily compared to other layer types (dense layers, recurrent layers, dropout layers, etc.). 
For both beginners and experts, it is difficult to properly build a model without knowing how the model \textit{see}s the dataset. 
Although there are tools for visualizing raw activations and high-level interpretation of CNN models, there is a gap in low-level explanations of the activation channels for any generic vision model.
In this paper, we present an image-based model visualization technique and its implementation, \toolname{}, which is a data-driven visualization tool that helps debug the models and datasets at the level of activation channels. 
We aim to keep the visualization code reproducible and extendable for general-purpose use. 
We believe the summarization methods and visualization techniques will help locate weaknesses and expose ways to improve them with minimal effort.

\setstretch{1}
\bibliographystyle{abbrv-doi-hyperref-narrow}
\bibliography{channel-explorer-main}

\begin{thebibliography}{10}
\renewcommand*{\sfdefault}{PTSansNarrow-TLF}

\bibitem{radviz}
Y.~Abraham and N.~Sauwen.
\newblock {\em Radviz: Project Multidimensional Data in 2D Space}, 2025.
\newblock R package version 0.9.3.

\bibitem{agarap_deep_2019}
A.~F. Agarap.
\newblock Deep {Learning} using {Rectified} {Linear} {Units} ({ReLU}).
\newblock {\em arXiv preprint}, 2019.
  \href{https://doi.org/10.48550/arXiv.1803.08375}
{doi: \textsf{%
10\hspace{.1pt}\discretionary{.}{%
}{.}\hspace{.4pt}48550\discretionary{/}{%
}{/}arXiv\hspace{.1pt}\discretionary{.}{%
}{.}\hspace{.4pt}1803\hspace{.1pt}\discretionary{.}{%
}{.}\hspace{.4pt}08375}}


\bibitem{histogramclassviz}
B.~Alsallakh, A.~Hanbury, H.~Hauser, S.~Miksch, and A.~Rauber.
\newblock Visual methods for analyzing probabilistic classification data.
\newblock {\em IEEE Transactions on Visualization and Computer Graphics},
  20(12):1703--1712, 2014. \href{https://doi.org/10.1109/TVCG.2014.2346660}
{doi: \textsf{%
10\hspace{.1pt}\discretionary{.}{%
}{.}\hspace{.4pt}1109\discretionary{/}{%
}{/}TVCG\hspace{.1pt}\discretionary{.}{%
}{.}\hspace{.4pt}2014\hspace{.1pt}\discretionary{.}{%
}{.}\hspace{.4pt}2346660}}


\bibitem{AlsallakhJourablooYe2018}
B.~Alsallakh, A.~Jourabloo, M.~Ye, X.~Liu, and L.~Ren.
\newblock Understanding the error structure as a key to regularize
  convolutional neural networks.
\newblock {\em Conference on Machine Learning and Systems}, 2018.

\bibitem{AyindePruning2018}
B.~Ayinde and J.~Zurada.
\newblock Building efficient convnets using redundant feature pruning.
\newblock {\em arXiv preprint}, 2018.
  \href{https://doi.org/10.48550/arXiv.1802.07653}
{doi: \textsf{%
10\hspace{.1pt}\discretionary{.}{%
}{.}\hspace{.4pt}48550\discretionary{/}{%
}{/}arXiv\hspace{.1pt}\discretionary{.}{%
}{.}\hspace{.4pt}1802\hspace{.1pt}\discretionary{.}{%
}{.}\hspace{.4pt}07653}}


\bibitem{BainTokarevKothari2021}
R.~Bain, M.~Tokarev, H.~Kothari, and R.~Damineni.
\newblock Lossplot: {A} better way to visualize loss landscapes.
\newblock {\em arXiv preprint}, 2021.
  \href{https://doi.org/10.48550/2111.15133}
{doi: \textsf{%
10\hspace{.1pt}\discretionary{.}{%
}{.}\hspace{.4pt}48550\discretionary{/}{%
}{/}2111\hspace{.1pt}\discretionary{.}{%
}{.}\hspace{.4pt}15133}}


\bibitem{DBLP:journals/corr/BauZKOT17}
D.~Bau, B.~Zhou, A.~Khosla, A.~Oliva, and A.~Torralba.
\newblock Network dissection: Quantifying interpretability of deep visual
  representations.
\newblock In {\em IEEE Computer Vision and Pattern Recognition (CVPR)}, pp.
  6541--6549, 2017. \href{https://doi.org/10.1109/CVPR.2017.354}
{doi: \textsf{%
10\hspace{.1pt}\discretionary{.}{%
}{.}\hspace{.4pt}1109\discretionary{/}{%
}{/}CVPR\hspace{.1pt}\discretionary{.}{%
}{.}\hspace{.4pt}2017\hspace{.1pt}\discretionary{.}{%
}{.}\hspace{.4pt}354}}


\bibitem{BelcherPrugel-BennettDasmahapatra2020}
D.~Belcher, A.~Prugel-Bennett, and S.~Dasmahapatra.
\newblock Generalisation and the geometry of class separability.
\newblock In {\em NeurIPS Workshop: Deep Learning through Information
  Geometry}, 2020.

\bibitem{bellgardt_immersive_2020}
M.~Bellgardt, C.~Scheiderer, and T.~W. Kuhlen.
\newblock An {Immersive} {Node}-{Link} {Visualization} of {Artificial} {Neural}
  {Networks} for {Machine} {Learning} {Experts}.
\newblock In {\em IEEE Conference on Artificial Intelligence and Virtual
  Reality (AIVR)}, pp. 33--36, 2020.
  \href{https://doi.org/10.1109/AIVR50618.2020.00015}
{doi: \textsf{%
10\hspace{.1pt}\discretionary{.}{%
}{.}\hspace{.4pt}1109\discretionary{/}{%
}{/}AIVR50618\hspace{.1pt}\discretionary{.}{%
}{.}\hspace{.4pt}2020\hspace{.1pt}\discretionary{.}{%
}{.}\hspace{.4pt}00015}}


\bibitem{nn_meaningful}
K.~Beyer, J.~Goldstein, R.~Ramakrishnan, and U.~Shaft.
\newblock When is ``nearest neighbor'' meaningful?
\newblock In {\em Database Theory --- ICDT}, pp. 217--235, 1999.
  \href{https://doi.org/10.1007/3-540-49257-7_15}
{doi: \textsf{%
10\hspace{.1pt}\discretionary{.}{%
}{.}\hspace{.4pt}1007\discretionary{/}{%
}{/}3\discretionary{%
}{-}{-}540\discretionary{%
}{-}{-}49257\discretionary{%
}{-}{-}7\_15}}


\bibitem{bianquiver}
J.~Bian.
\newblock Quiver.
\newblock \url{https://github.com/keplr-io/quiver}, 2016.

\bibitem{BilalJourablooYe2017}
A.~Bilal, A.~Jourabloo, M.~Ye, X.~Liu, and L.~Ren.
\newblock Do convolutional neural networks learn class hierarchy?
\newblock {\em IEEE Transactions on Visualization and Computer Graphics},
  24(1):152--162, 2017. \href{https://doi.org/10.1109/TVCG.2017.2744683}
{doi: \textsf{%
10\hspace{.1pt}\discretionary{.}{%
}{.}\hspace{.4pt}1109\discretionary{/}{%
}{/}TVCG\hspace{.1pt}\discretionary{.}{%
}{.}\hspace{.4pt}2017\hspace{.1pt}\discretionary{.}{%
}{.}\hspace{.4pt}2744683}}


\bibitem{carter2019activation}
S.~Carter, Z.~Armstrong, L.~Schubert, I.~Johnson, and C.~Olah.
\newblock Activation atlas.
\newblock {\em Distill}, 2019. \href{https://doi.org/10.23915/distill.00015}
{doi: \textsf{%
10\hspace{.1pt}\discretionary{.}{%
}{.}\hspace{.4pt}23915\discretionary{/}{%
}{/}distill\hspace{.1pt}\discretionary{.}{%
}{.}\hspace{.4pt}00015}}


\bibitem{barnhill_class_1974}
E.~Catmull and R.~Rom.
\newblock A class of local interpolating splines.
\newblock In {\em Computer Aided Geometric Design}, pp. 317--326. Academic
  Press, 1974. \href{https://doi.org/10.1016/B978-0-12-079050-0.50020-5}
{doi: \textsf{%
10\hspace{.1pt}\discretionary{.}{%
}{.}\hspace{.4pt}1016\discretionary{/}{%
}{/}B978\discretionary{%
}{-}{-}0\discretionary{%
}{-}{-}12\discretionary{%
}{-}{-}079050\discretionary{%
}{-}{-}0\hspace{.1pt}\discretionary{.}{%
}{.}\hspace{.4pt}50020\discretionary{%
}{-}{-}5}}


\bibitem{cho2024transformerexplainerinteractivelearning}
A.~Cho, G.~C. Kim, A.~Karpekov, A.~Helbling, Z.~J. Wang, S.~Lee, B.~Hoover, and
  D.~H. Chau.
\newblock Transformer explainer: Interactive learning of text-generative
  models.
\newblock {\em arXiv preprint}, 2024.
  \href{https://doi.org/10.48550/arXiv.2408.04619}
{doi: \textsf{%
10\hspace{.1pt}\discretionary{.}{%
}{.}\hspace{.4pt}48550\discretionary{/}{%
}{/}arXiv\hspace{.1pt}\discretionary{.}{%
}{.}\hspace{.4pt}2408\hspace{.1pt}\discretionary{.}{%
}{.}\hspace{.4pt}04619}}


\bibitem{revacnn}
S.~Chung, S.~Suh, C.~Park, K.~Kang, J.~Choo, and B.~C. Kwon.
\newblock {Revacnn}: {Real-Time} visual analytics for convolutional neural
  network.
\newblock {\em ACM SIGKDD Workshop on Interactive Data Exploration and
  Analytics (IDEA)}, 2016.

\bibitem{clevert_fast_2016}
D.-A. Clevert, T.~Unterthiner, and S.~Hochreiter.
\newblock Fast and {Accurate} {Deep} {Network} {Learning} by {Exponential}
  {Linear} {Units} ({ELUs}).
\newblock {\em arXiv preprint}, 2016.
  \href{https://doi.org/10.48550/arXiv.1511.07289}
{doi: \textsf{%
10\hspace{.1pt}\discretionary{.}{%
}{.}\hspace{.4pt}48550\discretionary{/}{%
}{/}arXiv\hspace{.1pt}\discretionary{.}{%
}{.}\hspace{.4pt}1511\hspace{.1pt}\discretionary{.}{%
}{.}\hspace{.4pt}07289}}


\bibitem{das2020bluff}
N.~Das, H.~Park, Z.~J. Wang, F.~Hohman, R.~Firstman, E.~Rogers, and D.~H.~P.
  Chau.
\newblock Bluff: Interactively deciphering adversarial attacks on deep neural
  networks.
\newblock In {\em IEEE Visualization Conference (VIS)}, pp. 271--275, 2020.
  \href{https://doi.org/10.1109/VIS47514.2020.00061}
{doi: \textsf{%
10\hspace{.1pt}\discretionary{.}{%
}{.}\hspace{.4pt}1109\discretionary{/}{%
}{/}VIS47514\hspace{.1pt}\discretionary{.}{%
}{.}\hspace{.4pt}2020\hspace{.1pt}\discretionary{.}{%
}{.}\hspace{.4pt}00061}}


\bibitem{10.1007/978-3-540-31843-9_17}
M.~Eiglsperger, M.~Siebenhaller, and M.~Kaufmann.
\newblock An efficient implementation of sugiyama's algorithm for layered graph
  drawing.
\newblock In {\em Graph Drawing}, pp. 155--166, 2005.
  \href{https://doi.org/10.1007/978-3-540-31843-9_17}
{doi: \textsf{%
10\hspace{.1pt}\discretionary{.}{%
}{.}\hspace{.4pt}1007\discretionary{/}{%
}{/}978\discretionary{%
}{-}{-}3\discretionary{%
}{-}{-}540\discretionary{%
}{-}{-}31843\discretionary{%
}{-}{-}9\_17}}


\bibitem{dumitruVisualizing2009}
D.~Erhan, Y.~Bengio, A.~Courville, and P.~Vincent.
\newblock Visualizing higher-layer features of a deep network.
\newblock {\em Technical Report, University of Montreal}, 2009.

\bibitem{DBLP:journals/corr/abs-1801-03454}
R.~Fong and A.~Vedaldi.
\newblock Net2vec: Quantifying and explaining how concepts are encoded by
  filters in deep neural networks.
\newblock In {\em IEEE Computer Vision and Pattern Recognition (CVPR)}, 2018.
  \href{https://doi.org/10.1109/CVPR.2018.00910}
{doi: \textsf{%
10\hspace{.1pt}\discretionary{.}{%
}{.}\hspace{.4pt}1109\discretionary{/}{%
}{/}CVPR\hspace{.1pt}\discretionary{.}{%
}{.}\hspace{.4pt}2018\hspace{.1pt}\discretionary{.}{%
}{.}\hspace{.4pt}00910}}


\bibitem{FraternaliMilaniTorres2022}
P.~Fraternali, F.~Milani, R.~N. Torres, and N.~Zangrando.
\newblock Black-box error diagnosis in deep neural networks: a survey of tools.
\newblock {\em arXiv preprint}, 2022.
  \href{https://doi.org/10.48550/2201.06444}
{doi: \textsf{%
10\hspace{.1pt}\discretionary{.}{%
}{.}\hspace{.4pt}48550\discretionary{/}{%
}{/}2201\hspace{.1pt}\discretionary{.}{%
}{.}\hspace{.4pt}06444}}


\bibitem{ace_ghorbani}
A.~Ghorbani, J.~Wexler, J.~Y. Zou, and B.~Kim.
\newblock Towards automatic concept-based explanations.
\newblock In {\em Advances in Neural Information Processing Systems}, pp.
  9273--9282, 2019.

\bibitem{GhoshBellingerCorizzo2024}
K.~Ghosh, C.~Bellinger, R.~Corizzo, P.~Branco, B.~Krawczyk, and N.~Japkowicz.
\newblock The class imbalance problem in deep learning.
\newblock {\em Machine Learning}, 113(7):4845--4901, 2024.
  \href{https://doi.org/10.1007/s10994-022-06268-8}
{doi: \textsf{%
10\hspace{.1pt}\discretionary{.}{%
}{.}\hspace{.4pt}1007\discretionary{/}{%
}{/}s10994\discretionary{%
}{-}{-}022\discretionary{%
}{-}{-}06268\discretionary{%
}{-}{-}8}}


\bibitem{GirshickDonahueDarrell2014}
R.~B. Girshick, J.~Donahue, T.~Darrell, and J.~Malik.
\newblock Rich feature hierarchies for accurate object detection and semantic
  segmentation.
\newblock In {\em IEEE Computer Vision and Pattern Recognition (CVPR)}, pp.
  580--587, 2014. \href{https://doi.org/10.1109/CVPR.2014.81}
{doi: \textsf{%
10\hspace{.1pt}\discretionary{.}{%
}{.}\hspace{.4pt}1109\discretionary{/}{%
}{/}CVPR\hspace{.1pt}\discretionary{.}{%
}{.}\hspace{.4pt}2014\hspace{.1pt}\discretionary{.}{%
}{.}\hspace{.4pt}81}}


\bibitem{goh2021multimodal}
G.~Goh, N.~Cammarata, C.~Voss, S.~Carter, M.~Petrov, L.~Schubert, A.~Radford,
  and C.~Olah.
\newblock Multimodal neurons in artificial neural networks.
\newblock {\em Distill}, 2021.
\newblock https://distill.pub/2021/multimodal-neurons.
  \href{https://doi.org/10.23915/distill.00030}
{doi: \textsf{%
10\hspace{.1pt}\discretionary{.}{%
}{.}\hspace{.4pt}23915\discretionary{/}{%
}{/}distill\hspace{.1pt}\discretionary{.}{%
}{.}\hspace{.4pt}00030}}


\bibitem{neo_conf_matrix}
J.~G{\"o}rtler, F.~Hohman, D.~Moritz, K.~Wongsuphasawat, D.~Ren, R.~Nair,
  M.~Kirchner, and K.~Patel.
\newblock Neo: Generalizing confusion matrix visualization to hierarchical and
  multi-output labels.
\newblock In {\em ACM SIGCHI Conference on Human Factors in Computing Systems},
  2022. \href{https://doi.org/10.1145/3491102.3501823}
{doi: \textsf{%
10\hspace{.1pt}\discretionary{.}{%
}{.}\hspace{.4pt}1145\discretionary{/}{%
}{/}3491102\hspace{.1pt}\discretionary{.}{%
}{.}\hspace{.4pt}3501823}}


\bibitem{GranziolWanGaripov2020}
D.~Granziol, X.~Wan, and T.~Garipov.
\newblock Deep curvature suite.
\newblock {\em arXiv preprint}, 2020.
  \href{https://doi.org/10.48550/arXiv.1912.09656}
{doi: \textsf{%
10\hspace{.1pt}\discretionary{.}{%
}{.}\hspace{.4pt}48550\discretionary{/}{%
}{/}arXiv\hspace{.1pt}\discretionary{.}{%
}{.}\hspace{.4pt}1912\hspace{.1pt}\discretionary{.}{%
}{.}\hspace{.4pt}09656}}


\bibitem{harley_interactive_2015}
A.~W. Harley.
\newblock An {Interactive} {Node}-{Link} {Visualization} of {Convolutional}
  {Neural} {Networks}.
\newblock In {\em International Symposium on Visual Computing (ISVC)}, pp.
  867--877, 2015. \href{https://doi.org/10.1007/978-3-319-27857-5_77}
{doi: \textsf{%
10\hspace{.1pt}\discretionary{.}{%
}{.}\hspace{.4pt}1007\discretionary{/}{%
}{/}978\discretionary{%
}{-}{-}3\discretionary{%
}{-}{-}319\discretionary{%
}{-}{-}27857\discretionary{%
}{-}{-}5\_77}}


\bibitem{he_delving_2015}
K.~He, X.~Zhang, S.~Ren, and J.~Sun.
\newblock Delving {Deep} into {Rectifiers}: {Surpassing} {Human}-{Level}
  {Performance} on {ImageNet} {Classification}.
\newblock {\em arXiv preprint}, 2015.
  \href{https://doi.org/10.48550/arXiv.1502.01852}
{doi: \textsf{%
10\hspace{.1pt}\discretionary{.}{%
}{.}\hspace{.4pt}48550\discretionary{/}{%
}{/}arXiv\hspace{.1pt}\discretionary{.}{%
}{.}\hspace{.4pt}1502\hspace{.1pt}\discretionary{.}{%
}{.}\hspace{.4pt}01852}}


\bibitem{confusionflow}
A.~Hinterreiter, P.~Ruch, H.~Stitz, M.~Ennemoser, J.~Bernard, H.~Strobelt, and
  M.~Streit.
\newblock Confusionflow: A model-agnostic visualization for temporal analysis
  of classifier confusion.
\newblock {\em IEEE Transactions on Visualization and Computer Graphics},
  28(2):1222--1236, 2022. \href{https://doi.org/10.1109/TVCG.2020.3012063}
{doi: \textsf{%
10\hspace{.1pt}\discretionary{.}{%
}{.}\hspace{.4pt}1109\discretionary{/}{%
}{/}TVCG\hspace{.1pt}\discretionary{.}{%
}{.}\hspace{.4pt}2020\hspace{.1pt}\discretionary{.}{%
}{.}\hspace{.4pt}3012063}}


\bibitem{DBLP:journals/corr/abs-2101-04141}
H.~Hoeiness, A.~Harstad, and G.~Friedland.
\newblock From tinkering to engineering: Measurements in tensorflow playground.
\newblock {\em arXiv preprint}, 2021.
  \href{https://doi.org/10.48550/2101.04141}
{doi: \textsf{%
10\hspace{.1pt}\discretionary{.}{%
}{.}\hspace{.4pt}48550\discretionary{/}{%
}{/}2101\hspace{.1pt}\discretionary{.}{%
}{.}\hspace{.4pt}04141}}


\bibitem{hohman2020summit}
F.~Hohman, H.~Park, C.~Robinson, and D.~H. Chau.
\newblock Summit: Scaling deep learning interpretability by visualizing
  activation and attribution summarizations.
\newblock {\em IEEE Transactions on Visualization and Computer Graphics}, 2020.
  \href{https://doi.org/10.1109/TVCG.2019.2934659}
{doi: \textsf{%
10\hspace{.1pt}\discretionary{.}{%
}{.}\hspace{.4pt}1109\discretionary{/}{%
}{/}TVCG\hspace{.1pt}\discretionary{.}{%
}{.}\hspace{.4pt}2019\hspace{.1pt}\discretionary{.}{%
}{.}\hspace{.4pt}2934659}}


\bibitem{HoroiHuangWolf2021}
S.~Horoi, J.~Huang, G.~Wolf, and S.~Krishnaswamy.
\newblock Visualizing high-dimensional trajectories on the loss-landscape of
  {ANNs}.
\newblock {\em OpenReview}, 2021.
\newblock https://openreview.net/forum?id=uhiF-dV99ir.

\bibitem{HuYangYi2015}
G.~Hu, Y.~Yang, D.~Yi, J.~Kittler, W.~J. Christmas, S.~Z. Li, and T.~M.
  Hospedales.
\newblock When face recognition meets with deep learning: an evaluation of
  convolutional neural networks for face recognition.
\newblock {\em IEEE International Conference on Computer Vision Workshop
  (ICCVW)}, 2015. \href{https://doi.org/10.1109/ICCVW.2015.58}
{doi: \textsf{%
10\hspace{.1pt}\discretionary{.}{%
}{.}\hspace{.4pt}1109\discretionary{/}{%
}{/}ICCVW\hspace{.1pt}\discretionary{.}{%
}{.}\hspace{.4pt}2015\hspace{.1pt}\discretionary{.}{%
}{.}\hspace{.4pt}58}}


\bibitem{conceptexplainer}
J.~Huang, A.~Mishra, B.~C. Kwon, and C.~Bryan.
\newblock Conceptexplainer: Interactive explanation for deep neural networks
  from a concept perspective.
\newblock {\em IEEE Transactions on Visualization and Computer Graphics},
  29(1):831--841, 2023. \href{https://doi.org/10.1109/TVCG.2022.3209384}
{doi: \textsf{%
10\hspace{.1pt}\discretionary{.}{%
}{.}\hspace{.4pt}1109\discretionary{/}{%
}{/}TVCG\hspace{.1pt}\discretionary{.}{%
}{.}\hspace{.4pt}2022\hspace{.1pt}\discretionary{.}{%
}{.}\hspace{.4pt}3209384}}


\bibitem{jaccard1902distribution}
P.~Jaccard.
\newblock {\'E}tude comparative de la distribution florale dans une portion des
  alpes et des jura.
\newblock {\em Bull Soc Vaudoise Sci Nat}, 37:547--579, 1901.

\bibitem{jacobs_cosaliency_2010}
D.~E. Jacobs, D.~B. Goldman, and E.~Shechtman.
\newblock Cosaliency: where people look when comparing images.
\newblock In {\em ACM Symposium on User Interface Software and Technology
  (UIST)}, pp. 219--228, 2010. \href{https://doi.org/10.1145/1866029.1866066}
{doi: \textsf{%
10\hspace{.1pt}\discretionary{.}{%
}{.}\hspace{.4pt}1145\discretionary{/}{%
}{/}1866029\hspace{.1pt}\discretionary{.}{%
}{.}\hspace{.4pt}1866066}}


\bibitem{10308641}
H.~Jeon, G.~J. Quadri, H.~Lee, P.~Rosen, D.~A. Szafir, and J.~Seo.
\newblock Clams: A cluster ambiguity measure for estimating perceptual
  variability in visual clustering.
\newblock {\em IEEE Transactions on Visualization and Computer Graphics},
  30(1):770--780, 2024. \href{https://doi.org/10.1109/TVCG.2023.3327201}
{doi: \textsf{%
10\hspace{.1pt}\discretionary{.}{%
}{.}\hspace{.4pt}1109\discretionary{/}{%
}{/}TVCG\hspace{.1pt}\discretionary{.}{%
}{.}\hspace{.4pt}2023\hspace{.1pt}\discretionary{.}{%
}{.}\hspace{.4pt}3327201}}


\bibitem{DBLP:journals/corr/KahngAKC17}
M.~Kahng, P.~Y. Andrews, A.~Kalro, and D.~H. Chau.
\newblock Activis: Visual exploration of industry-scale deep neural network
  models.
\newblock {\em IEEE Transactions on Visualization and Computer Graphics}, 2017.
  \href{https://doi.org/10.1109/TVCG.2017.2744718}
{doi: \textsf{%
10\hspace{.1pt}\discretionary{.}{%
}{.}\hspace{.4pt}1109\discretionary{/}{%
}{/}TVCG\hspace{.1pt}\discretionary{.}{%
}{.}\hspace{.4pt}2017\hspace{.1pt}\discretionary{.}{%
}{.}\hspace{.4pt}2744718}}


\bibitem{Kahng2018GANLU}
M.~Kahng, N.~Thorat, D.~H. Chau, F.~B. Vi{\'e}gas, and M.~Wattenberg.
\newblock Gan lab: Understanding complex deep generative models using
  interactive visual experimentation.
\newblock {\em IEEE Transactions on Visualization and Computer Graphics},
  25:310--320, 2018. \href{https://doi.org/10.1109/TVCG.2018.2864500}
{doi: \textsf{%
10\hspace{.1pt}\discretionary{.}{%
}{.}\hspace{.4pt}1109\discretionary{/}{%
}{/}TVCG\hspace{.1pt}\discretionary{.}{%
}{.}\hspace{.4pt}2018\hspace{.1pt}\discretionary{.}{%
}{.}\hspace{.4pt}2864500}}


\bibitem{manimatrix}
A.~Kapoor, B.~Lee, D.~Tan, and E.~Horvitz.
\newblock Interactive optimization for steering machine classification.
\newblock In {\em ACM SIGCHI Conference on Human Factors in Computing Systems},
  pp. 1343--1352. Association for Computing Machinery, 2010.
  \href{https://doi.org/10.1145/1753326.1753529}
{doi: \textsf{%
10\hspace{.1pt}\discretionary{.}{%
}{.}\hspace{.4pt}1145\discretionary{/}{%
}{/}1753326\hspace{.1pt}\discretionary{.}{%
}{.}\hspace{.4pt}1753529}}


\bibitem{pmlr-v80-kim18d}
B.~Kim, M.~Wattenberg, J.~Gilmer, C.~Cai, J.~Wexler, F.~Viegas, and R.~Sayres.
\newblock Interpretability beyond feature attribution: Quantitative testing
  with concept activation vectors ({TCAV}).
\newblock In {\em International Conference on Machine Learning}, vol.~80 of
  {\em Proceedings of Machine Learning Research}, pp. 2668--2677, 2018.

\bibitem{klambauer_self-normalizing_2017}
G.~Klambauer, T.~Unterthiner, A.~Mayr, and S.~Hochreiter.
\newblock Self-{Normalizing} {Neural} {Networks}.
\newblock {\em arXiv preprint}, 2017.
  \href{https://doi.org/10.48550/arXiv.1706.02515}
{doi: \textsf{%
10\hspace{.1pt}\discretionary{.}{%
}{.}\hspace{.4pt}48550\discretionary{/}{%
}{/}arXiv\hspace{.1pt}\discretionary{.}{%
}{.}\hspace{.4pt}1706\hspace{.1pt}\discretionary{.}{%
}{.}\hspace{.4pt}02515}}


\bibitem{vitmodel}
A.~Kolesnikov, A.~Dosovitskiy, D.~Weissenborn, G.~Heigold, J.~Uszkoreit,
  L.~Beyer, M.~Minderer, M.~Dehghani, N.~Houlsby, S.~Gelly, T.~Unterthiner, and
  X.~Zhai.
\newblock An image is worth 16x16 words: Transformers for image recognition at
  scale.
\newblock 2021.

\bibitem{KrizhevskySutskeverHinton2017}
A.~Krizhevsky, I.~Sutskever, and G.~E. Hinton.
\newblock Imagenet classification with deep convolutional neural networks.
\newblock {\em Communications of the {ACM}}, 60(6):84--90, 2017.
  \href{https://doi.org/10.1145/3065386}
{doi: \textsf{%
10\hspace{.1pt}\discretionary{.}{%
}{.}\hspace{.4pt}1145\discretionary{/}{%
}{/}3065386}}


\bibitem{kruskal_multidimensional_1964}
J.~B. Kruskal.
\newblock Multidimensional scaling by optimizing goodness of fit to a nonmetric
  hypothesis.
\newblock {\em Psychometrika}, 29(1):1--27, 1964.
  \href{https://doi.org/10.1007/BF02289565}
{doi: \textsf{%
10\hspace{.1pt}\discretionary{.}{%
}{.}\hspace{.4pt}1007\discretionary{/}{%
}{/}BF02289565}}


\bibitem{srresnet2017}
C.~Ledig, L.~Theis, F.~Husz{\'a}r, J.~Caballero, A.~Cunningham, A.~Acosta,
  A.~Aitken, A.~Tejani, J.~Totz, Z.~Wang, and W.~Shi.
\newblock Photo-realistic single image super-resolution using a generative
  adversarial network.
\newblock In {\em IEEE Computer Vision and Pattern Recognition (CVPR)}, pp.
  105--114, 2017. \href{https://doi.org/10.1109/CVPR.2017.19}
{doi: \textsf{%
10\hspace{.1pt}\discretionary{.}{%
}{.}\hspace{.4pt}1109\discretionary{/}{%
}{/}CVPR\hspace{.1pt}\discretionary{.}{%
}{.}\hspace{.4pt}2017\hspace{.1pt}\discretionary{.}{%
}{.}\hspace{.4pt}19}}


\bibitem{lee2024diffusionexplainervisualexplanation}
S.~Lee, B.~Hoover, H.~Strobelt, Z.~J. Wang, S.~Peng, A.~Wright, K.~Li, H.~Park,
  H.~Yang, and D.~H. Chau.
\newblock Diffusion explainer: Visual explanation for text-to-image stable
  diffusion.
\newblock {\em arXiv preprint}, 2024.
  \href{https://doi.org/10.48550/arXiv.2305.03509}
{doi: \textsf{%
10\hspace{.1pt}\discretionary{.}{%
}{.}\hspace{.4pt}48550\discretionary{/}{%
}{/}arXiv\hspace{.1pt}\discretionary{.}{%
}{.}\hspace{.4pt}2305\hspace{.1pt}\discretionary{.}{%
}{.}\hspace{.4pt}03509}}


\bibitem{samuel_visualizing_2018}
S.~P. Leeman-Munk, S.~Sethi, C.~G. Healey, S.~Nie, K.~Padia, R.~Devarajan,
  D.~J. Caira, J.~R. Benson, J.~A. Cox, L.~E. Lewis, et~al.
\newblock Visualizing convolutional neural networks, 2019.
\newblock US Patent 10,192,001.

\bibitem{deeptracker}
D.~Liu, W.~Cui, K.~Jin, Y.~Guo, and H.~Qu.
\newblock Deeptracker: Visualizing the training process of convolutional neural
  networks.
\newblock {\em {ACM} Trans. Intell. Syst. Technol.}, 10(1):6:1--6:25, 2019.
  \href{https://doi.org/10.1145/3200489}
{doi: \textsf{%
10\hspace{.1pt}\discretionary{.}{%
}{.}\hspace{.4pt}1145\discretionary{/}{%
}{/}3200489}}


\bibitem{DBLP:journals/corr/abs-1810-03913}
M.~Liu, S.~Liu, H.~Su, K.~Cao, and J.~Zhu.
\newblock Analyzing the noise robustness of deep neural networks.
\newblock {\em arXiv preprint}, 2018.
  \href{https://doi.org/10.1109/TVCG.2020.2969185}
{doi: \textsf{%
10\hspace{.1pt}\discretionary{.}{%
}{.}\hspace{.4pt}1109\discretionary{/}{%
}{/}TVCG\hspace{.1pt}\discretionary{.}{%
}{.}\hspace{.4pt}2020\hspace{.1pt}\discretionary{.}{%
}{.}\hspace{.4pt}2969185}}


\bibitem{liu2022convnet}
Z.~Liu, H.~Mao, C.-Y. Wu, C.~Feichtenhofer, T.~Darrell, and S.~Xie.
\newblock A convnet for the 2020s.
\newblock {\em IEEE Computer Vision and Pattern Recognition (CVPR)}, 2022.
  \href{https://doi.org/10.1109/CVPR52688.2022.01167}
{doi: \textsf{%
10\hspace{.1pt}\discretionary{.}{%
}{.}\hspace{.4pt}1109\discretionary{/}{%
}{/}CVPR52688\hspace{.1pt}\discretionary{.}{%
}{.}\hspace{.4pt}2022\hspace{.1pt}\discretionary{.}{%
}{.}\hspace{.4pt}01167}}


\bibitem{LuoWongKankanhalli2020}
Y.~Luo, Y.~Wong, M.~Kankanhalli, and Q.~Zhao.
\newblock $\mathcal{G}$ -softmax: Improving intraclass compactness and
  interclass separability of features.
\newblock {\em IEEE Transactions on Neural Networks and Learning Systems},
  31(2):685--699, 2020. \href{https://doi.org/10.1109/TNNLS.2019.2909737}
{doi: \textsf{%
10\hspace{.1pt}\discretionary{.}{%
}{.}\hspace{.4pt}1109\discretionary{/}{%
}{/}TNNLS\hspace{.1pt}\discretionary{.}{%
}{.}\hspace{.4pt}2019\hspace{.1pt}\discretionary{.}{%
}{.}\hspace{.4pt}2909737}}


\bibitem{maaten_visualizing_2008}
L.~v.~d. Maaten and G.~Hinton.
\newblock Visualizing {Data} using t-{SNE}.
\newblock {\em Journal of Machine Learning Research}, 9(86):2579--2605, 2008.

\bibitem{mackiewicz_principal_1993}
A.~Ma{\'c}kiewicz and W.~Ratajczak.
\newblock Principal components analysis ({PCA}).
\newblock {\em Computers \& Geosciences}, 19(3):303--342, 1993.
  \href{https://doi.org/10.1016/0098-3004(93)90090-R}
{doi: \textsf{%
10\hspace{.1pt}\discretionary{.}{%
}{.}\hspace{.4pt}1016\discretionary{/}{%
}{/}0098\discretionary{%
}{-}{-}3004\discretionary{%
}{(}{(}93\discretionary{)}{%
}{)}90090\discretionary{%
}{-}{-}R}}


\bibitem{mcinnes_umap_2020}
L.~McInnes, J.~Healy, and J.~Melville.
\newblock {UMAP}: {Uniform} {Manifold} {Approximation} and {Projection} for
  {Dimension} {Reduction}.
\newblock {\em arXiv preprint}, 2020.
  \href{https://doi.org/10.48550/arXiv.1802.03426}
{doi: \textsf{%
10\hspace{.1pt}\discretionary{.}{%
}{.}\hspace{.4pt}48550\discretionary{/}{%
}{/}arXiv\hspace{.1pt}\discretionary{.}{%
}{.}\hspace{.4pt}1802\hspace{.1pt}\discretionary{.}{%
}{.}\hspace{.4pt}03426}}


\bibitem{molnar2022}
C.~Molnar.
\newblock {\em Interpretable Machine Learning}.
\newblock Web, 2 ed., 2022.

\bibitem{google_inceptionism_2015}
A.~Mordvintsev, C.~Olah, and M.~Tyka.
\newblock Inceptionism: Going deeper into neural networks, 2015.

\bibitem{DBLP:journals/corr/NguyenYBDC16}
A.~Nguyen, J.~Yosinski, Y.~Bengio, A.~Dosovitskiy, and J.~Clune.
\newblock Plug {\&} play generative networks: Conditional iterative generation
  of images in latent space.
\newblock In {\em IEEE Computer Vision and Pattern Recognition (CVPR)}, 2016.
  \href{https://doi.org/10.1109/CVPR.2017.374}
{doi: \textsf{%
10\hspace{.1pt}\discretionary{.}{%
}{.}\hspace{.4pt}1109\discretionary{/}{%
}{/}CVPR\hspace{.1pt}\discretionary{.}{%
}{.}\hspace{.4pt}2017\hspace{.1pt}\discretionary{.}{%
}{.}\hspace{.4pt}374}}


\bibitem{olah2017feature}
C.~Olah, A.~Mordvintsev, and L.~Schubert.
\newblock Feature visualization.
\newblock {\em Distill}, 2017.
\newblock https://distill.pub/2017/feature-visualization.
  \href{https://doi.org/10.23915/distill.00007}
{doi: \textsf{%
10\hspace{.1pt}\discretionary{.}{%
}{.}\hspace{.4pt}23915\discretionary{/}{%
}{/}distill\hspace{.1pt}\discretionary{.}{%
}{.}\hspace{.4pt}00007}}


\bibitem{OlahSatyanarayanJohnson2018}
C.~Olah, A.~Satyanarayan, I.~Johnson, S.~Carter, L.~Schubert, K.~Ye, and
  A.~Mordvintsev.
\newblock The building blocks of interpretability.
\newblock {\em Distill}, 2018. \href{https://doi.org/10.23915/distill.00010}
{doi: \textsf{%
10\hspace{.1pt}\discretionary{.}{%
}{.}\hspace{.4pt}23915\discretionary{/}{%
}{/}distill\hspace{.1pt}\discretionary{.}{%
}{.}\hspace{.4pt}00010}}


\bibitem{openai_microscope}
{OpenAI} {Microscope}.

\bibitem{otsu}
N.~Otsu.
\newblock A threshold selection method from gray-level histograms.
\newblock {\em IEEE Transactions on Systems, Man, and Cybernetics},
  9(1):62--66, 1979. \href{https://doi.org/10.1109/TSMC.1979.4310076}
{doi: \textsf{%
10\hspace{.1pt}\discretionary{.}{%
}{.}\hspace{.4pt}1109\discretionary{/}{%
}{/}TSMC\hspace{.1pt}\discretionary{.}{%
}{.}\hspace{.4pt}1979\hspace{.1pt}\discretionary{.}{%
}{.}\hspace{.4pt}4310076}}


\bibitem{xmeans}
D.~Pelleg and A.~W. Moore.
\newblock X-means: Extending k-means with efficient estimation of the number of
  clusters.
\newblock In {\em International Conference on Machine Learning}, ICML,  8
  pages, pp. 727--734, 2000.

\bibitem{HSNE_Pezzotti}
N.~Pezzotti, T.~H{\"o}llt, B.~Lelieveldt, E.~Eisemann, and A.~Vilanova.
\newblock Hierarchical stochastic neighbor embedding.
\newblock {\em Computer Graphics Forum}, 35(3):21--30, 2016.
  \href{https://doi.org/10.1111/cgf.12878}
{doi: \textsf{%
10\hspace{.1pt}\discretionary{.}{%
}{.}\hspace{.4pt}1111\discretionary{/}{%
}{/}cgf\hspace{.1pt}\discretionary{.}{%
}{.}\hspace{.4pt}12878}}


\bibitem{pezzotti2017deepeyes}
N.~Pezzotti, T.~H{\"o}llt, J.~Van~Gemert, B.~P. Lelieveldt, E.~Eisemann, and
  A.~Vilanova.
\newblock Deepeyes: Progressive visual analytics for designing deep neural
  networks.
\newblock {\em IEEE Transactions on Visualization and Computer Graphics},
  24(1):98--108, 2017. \href{https://doi.org/10.1109/TVCG.2017.2744358}
{doi: \textsf{%
10\hspace{.1pt}\discretionary{.}{%
}{.}\hspace{.4pt}1109\discretionary{/}{%
}{/}TVCG\hspace{.1pt}\discretionary{.}{%
}{.}\hspace{.4pt}2017\hspace{.1pt}\discretionary{.}{%
}{.}\hspace{.4pt}2744358}}


\bibitem{instanceflow}
M.~P{\"u}hringer, A.~Hinterreiter, and M.~Streit.
\newblock Instanceflow: Visualizing the evolution of classifier confusion at
  the instance level.
\newblock In {\em IEEE Visualization Conference -- Short Papers}, 2020.
  \href{https://doi.org/10.1109/VIS47514.2020.00065}
{doi: \textsf{%
10\hspace{.1pt}\discretionary{.}{%
}{.}\hspace{.4pt}1109\discretionary{/}{%
}{/}VIS47514\hspace{.1pt}\discretionary{.}{%
}{.}\hspace{.4pt}2020\hspace{.1pt}\discretionary{.}{%
}{.}\hspace{.4pt}00065}}


\bibitem{pypi}
Python package index - pypi.
\newblock https://pypi.org/.

\bibitem{DBLP:journals/corr/abs-2103-00020}
A.~Radford, J.~W. Kim, C.~Hallacy, A.~Ramesh, G.~Goh, S.~Agarwal, G.~Sastry,
  A.~Askell, P.~Mishkin, J.~Clark, G.~Krueger, and I.~Sutskever.
\newblock Learning transferable visual models from natural language
  supervision.
\newblock {\em CoRR}, abs/2103.00020, 2021.

\bibitem{rathore_topoact_2021}
A.~Rathore, N.~Chalapathi, S.~Palande, and B.~Wang.
\newblock {TopoAct}: {Visually} {Exploring} the {Shape} of {Activations} in
  {Deep} {Learning}.
\newblock {\em Computer Graphics Forum}, 40(1):382--397, 2021.
  \href{https://doi.org/10.1111/cgf.14195}
{doi: \textsf{%
10\hspace{.1pt}\discretionary{.}{%
}{.}\hspace{.4pt}1111\discretionary{/}{%
}{/}cgf\hspace{.1pt}\discretionary{.}{%
}{.}\hspace{.4pt}14195}}


\bibitem{yolov3}
J.~Redmon and A.~Farhadi.
\newblock Yolov3: An incremental improvement.
\newblock {\em arXiv preprint}, 2018.

\bibitem{ren2016squares}
D.~Ren, S.~Amershi, B.~Lee, J.~Suh, and J.~Williams.
\newblock Squares: Supporting interactive performance analysis for multiclass
  classifiers.
\newblock In {\em IEEE VAST}, August 2016.
  \href{https://doi.org/10.1109/TVCG.2016.2598828}
{doi: \textsf{%
10\hspace{.1pt}\discretionary{.}{%
}{.}\hspace{.4pt}1109\discretionary{/}{%
}{/}TVCG\hspace{.1pt}\discretionary{.}{%
}{.}\hspace{.4pt}2016\hspace{.1pt}\discretionary{.}{%
}{.}\hspace{.4pt}2598828}}


\bibitem{rosenfeld_totally_2018}
A.~Rosenfeld, M.~D. Solbach, and J.~K. Tsotsos.
\newblock Totally {Looks} {Like} - {How} {Humans} {Compare}, {Compared} to
  {Machines}.
\newblock In {\em IEEE Computer Vision and Pattern Recognition Workshops}, pp.
  1961--1964, 2018. \href{https://doi.org/10.1007/978-3-030-20887-5_18}
{doi: \textsf{%
10\hspace{.1pt}\discretionary{.}{%
}{.}\hspace{.4pt}1007\discretionary{/}{%
}{/}978\discretionary{%
}{-}{-}3\discretionary{%
}{-}{-}030\discretionary{%
}{-}{-}20887\discretionary{%
}{-}{-}5\_18}}


\bibitem{schorr_neuroscope_2021}
C.~Schorr, P.~Goodarzi, F.~Chen, and T.~Dahmen.
\newblock Neuroscope: {An} {Explainable} {AI} {Toolbox} for {Semantic}
  {Segmentation} and {Image} {Classification} of {Convolutional} {Neural}
  {Nets}.
\newblock {\em Applied Sciences}, 11(5):2199, 2021.
  \href{https://doi.org/10.3390/app11052199}
{doi: \textsf{%
10\hspace{.1pt}\discretionary{.}{%
}{.}\hspace{.4pt}3390\discretionary{/}{%
}{/}app11052199}}


\bibitem{seifert_visualizations_2017}
C.~Seifert, A.~Aamir, A.~Balagopalan, D.~Jain, A.~Sharma, S.~Grottel, and
  S.~Gumhold.
\newblock Visualizations of {Deep} {Neural} {Networks} in {Computer} {Vision}:
  {A} {Survey}.
\newblock In {\em Transparent {Data} {Mining} for {Big} and {Small} {Data}},
  pp. 123--144, 2017. \href{https://doi.org/10.1007/978-3-319-54024-5_6}
{doi: \textsf{%
10\hspace{.1pt}\discretionary{.}{%
}{.}\hspace{.4pt}1007\discretionary{/}{%
}{/}978\discretionary{%
}{-}{-}3\discretionary{%
}{-}{-}319\discretionary{%
}{-}{-}54024\discretionary{%
}{-}{-}5\_6}}


\bibitem{shneiderman_eyes_1996}
B.~Shneiderman.
\newblock The eyes have it: a task by data type taxonomy for information
  visualizations.
\newblock In {\em IEEE Symposium on Visual Languages}, pp. 336--343, 1996.
  \href{https://doi.org/10.1109/VL.1996.545307}
{doi: \textsf{%
10\hspace{.1pt}\discretionary{.}{%
}{.}\hspace{.4pt}1109\discretionary{/}{%
}{/}VL\hspace{.1pt}\discretionary{.}{%
}{.}\hspace{.4pt}1996\hspace{.1pt}\discretionary{.}{%
}{.}\hspace{.4pt}545307}}


\bibitem{SultanaSufianDutta2020}
F.~Sultana, A.~Sufian, and P.~Dutta.
\newblock Evolution of image segmentation using deep convolutional neural
  network: {A} survey.
\newblock {\em Knowledge-Based Systems}, 2020.
  \href{https://doi.org/10.1016/j.knosys.2020.106062}
{doi: \textsf{%
10\hspace{.1pt}\discretionary{.}{%
}{.}\hspace{.4pt}1016\discretionary{/}{%
}{/}j\hspace{.1pt}\discretionary{.}{%
}{.}\hspace{.4pt}knosys\hspace{.1pt}\discretionary{.}{%
}{.}\hspace{.4pt}2020\hspace{.1pt}\discretionary{.}{%
}{.}\hspace{.4pt}106062}}


\bibitem{ensemblematrix}
J.~Talbot, B.~Lee, A.~Kapoor, and D.~S. Tan.
\newblock Ensemblematrix: interactive visualization to support machine learning
  with multiple classifiers.
\newblock In {\em ACM SIGCHI Conference on Human Factors in Computing Systems},
  pp. 1283--1292, 2009. \href{https://doi.org/10.1145/1518701.1518895}
{doi: \textsf{%
10\hspace{.1pt}\discretionary{.}{%
}{.}\hspace{.4pt}1145\discretionary{/}{%
}{/}1518701\hspace{.1pt}\discretionary{.}{%
}{.}\hspace{.4pt}1518895}}


\bibitem{AndrejT-SNE}
t-sne visualization of cnn codes.
\newblock \url{https://cs.stanford.edu/people/karpathy/cnnembed/}.

\bibitem{DBLP:journals/corr/TzengHZSD14}
E.~Tzeng, J.~Hoffman, N.~Zhang, K.~Saenko, and T.~Darrell.
\newblock Deep domain confusion: Maximizing for domain invariance.
\newblock {\em arXiv preprint}, 2014.
  \href{https://doi.org/10.48550/arXiv.1412.3474}
{doi: \textsf{%
10\hspace{.1pt}\discretionary{.}{%
}{.}\hspace{.4pt}48550\discretionary{/}{%
}{/}arXiv\hspace{.1pt}\discretionary{.}{%
}{.}\hspace{.4pt}1412\hspace{.1pt}\discretionary{.}{%
}{.}\hspace{.4pt}3474}}


\bibitem{dqnviz}
J.~Wang, L.~Gou, H.~Shen, and H.~Yang.
\newblock Dqnviz: {A} visual analytics approach to understand deep q-networks.
\newblock {\em IEEE Transactions on Visualization and Computer Graphics},
  25(1):288--298, 2019. \href{https://doi.org/10.1109/TVCG.2018.2864504}
{doi: \textsf{%
10\hspace{.1pt}\discretionary{.}{%
}{.}\hspace{.4pt}1109\discretionary{/}{%
}{/}TVCG\hspace{.1pt}\discretionary{.}{%
}{.}\hspace{.4pt}2018\hspace{.1pt}\discretionary{.}{%
}{.}\hspace{.4pt}2864504}}


\bibitem{Wang2020CNNEL}
Z.~J. Wang, R.~Turko, O.~Shaikh, H.~Park, N.~Das, F.~Hohman, M.~Kahng, and
  D.~H. Chau.
\newblock Cnn explainer: Learning convolutional neural networks with
  interactive visualization.
\newblock {\em IEEE Transactions on Visualization and Computer Graphics},
  27:1396--1406, 2020. \href{https://doi.org/10.1109/TVCG.2020.3030418}
{doi: \textsf{%
10\hspace{.1pt}\discretionary{.}{%
}{.}\hspace{.4pt}1109\discretionary{/}{%
}{/}TVCG\hspace{.1pt}\discretionary{.}{%
}{.}\hspace{.4pt}2020\hspace{.1pt}\discretionary{.}{%
}{.}\hspace{.4pt}3030418}}


\bibitem{XueZhangJiang2023}
L.~Xue, X.~Zhang, W.~Jiang, K.~Huo, and Q.~Shen.
\newblock A classification performance evaluation measure considering data
  separability.
\newblock In {\em Artificial Neural Networks and Machine Learning -- ICANN},
  pp. 1--13, 2023. \href{https://doi.org/10.1007/978-3-031-44207-0_1}
{doi: \textsf{%
10\hspace{.1pt}\discretionary{.}{%
}{.}\hspace{.4pt}1007\discretionary{/}{%
}{/}978\discretionary{%
}{-}{-}3\discretionary{%
}{-}{-}031\discretionary{%
}{-}{-}44207\discretionary{%
}{-}{-}0\_1}}


\bibitem{DBLP:journals/corr/YosinskiCNFL15}
J.~Yosinski, J.~Clune, A.~M. Nguyen, T.~J. Fuchs, and H.~Lipson.
\newblock Understanding neural networks through deep visualization.
\newblock {\em arXiv preprint}, 2015.
  \href{https://doi.org/10.48550/arXiv.1506.06579}
{doi: \textsf{%
10\hspace{.1pt}\discretionary{.}{%
}{.}\hspace{.4pt}48550\discretionary{/}{%
}{/}arXiv\hspace{.1pt}\discretionary{.}{%
}{.}\hspace{.4pt}1506\hspace{.1pt}\discretionary{.}{%
}{.}\hspace{.4pt}06579}}


\bibitem{YuZhu2020}
T.~Yu and H.~Zhu.
\newblock Hyper-parameter optimization: {A} review of algorithms and
  applications.
\newblock {\em arXiv preprint}, 2020.
  \href{https://doi.org/10.48550/arXiv.2003.05689}
{doi: \textsf{%
10\hspace{.1pt}\discretionary{.}{%
}{.}\hspace{.4pt}48550\discretionary{/}{%
}{/}arXiv\hspace{.1pt}\discretionary{.}{%
}{.}\hspace{.4pt}2003\hspace{.1pt}\discretionary{.}{%
}{.}\hspace{.4pt}05689}}


\bibitem{cnncomparator}
H.~Zeng, H.~Haleem, X.~Plantaz, N.~Cao, and H.~Qu.
\newblock Cnncomparator: Comparative analytics of convolutional neural
  networks.
\newblock {\em arXiv preprint}, 2017.

\end{thebibliography}

\end{document}